\documentclass[sn-mathphys,Numbered]{sn-jnl}

\usepackage{graphicx}%
\usepackage{multirow}%
\usepackage{amsmath,amssymb,amsfonts}%
\usepackage{amsthm}%
\usepackage{mathrsfs}%
\usepackage[title]{appendix}%
\usepackage{xcolor}%
\usepackage{textcomp}%
\usepackage{manyfoot}%
\usepackage{booktabs}%
\usepackage{algorithm}%
\usepackage{algorithmicx}%
\usepackage{algpseudocode}%
\usepackage{listings}%
\usepackage{bbold}%
\usepackage{comment}

\newcommand{\mat}[1]{\boldsymbol{#1}}

\begin{document}

\title[Article Title]{Scalable quantum detector tomography by high-performance computing}


\author*[1,2]{\fnm{Timon} \sur{Schapeler}}\email{timon.schapeler@uni-paderborn.de}
\equalcont{These authors contributed equally to this work.}

\author[3]{\fnm{Robert} \sur{Schade}}\email{robert.schade@uni-paderborn.de}
\equalcont{These authors contributed equally to this work.}

\author[3,4]{\fnm{Michael} \sur{Lass}}
\author[3,4]{\fnm{Christian} \sur{Plessl}}
\author[1,2]{\fnm{Tim J.} \sur{Bartley}}

\affil[1]{Department of Physics, Paderborn University, Warburger Str. 100, 33098 Paderborn, Germany}
\affil[2]{Institute for Photonic Quantum Systems (PhoQS), Paderborn University, Warburger Str. 100, 33098 Paderborn, Germany}

\affil[3]{Paderborn Center for Parallel Computing, Paderborn University, Warburger Str. 100, 33098 Paderborn, Germany}
\affil[4]{Department for Computer Science, Paderborn University, Warburger Str. 100, 33098 Paderborn, Germany}

\abstract{At large scales, quantum systems may become advantageous over their classical counterparts at performing certain tasks. Developing tools to analyse these systems at the relevant scales, in a manner consistent with quantum mechanics, is therefore critical to benchmarking performance and characterising their operation. While classical computational approaches cannot perform like-for-like computations of quantum systems beyond a certain scale, classical high-performance computing (HPC) may nevertheless be useful for precisely these characterisation and certification tasks. By developing open-source customised algorithms using high-performance computing, we perform quantum tomography on a megascale quantum photonic detector covering a Hilbert space of $10^6$. This requires finding $10^8$ elements of the matrix corresponding to the positive operator valued measure (POVM), the quantum description of the detector, and is achieved in minutes of computation time.  Moreover, by exploiting the structure of the problem, we achieve highly efficient parallel scaling, paving the way for quantum objects up to a system size of $10^{12}$ elements to be reconstructed using this method. In general, this shows that a consistent quantum mechanical description of quantum phenomena is applicable at everyday scales. More concretely, this enables the reconstruction of large-scale quantum sources, processes and detectors used in computation and sampling tasks, which may be necessary to prove their nonclassical character or quantum computational advantage.}




\maketitle
Photonic quantum computing paradigms are built around large scale generation, manipulation and measurement of quantum light. At sufficient scale, the computations these devices perform cannot be verified by conventional computing.
Pertinent examples are quantum simulators~\cite{aspuru2012photonic} and Boson sampling~\cite{aaronson2011the}, where in the latter beyond a certain system size, the process of sampling from the output distribution of a nonclassical input state in a photonic circuit implementing a random unitary matrix is a task which is computationally easier for a photonic quantum processor. Nevertheless, the same photonic processor, under illumination from ``classical'' light, is computationally ``easy'' to compute using classical approaches. Since this classical light is tomographically complete, one can use techniques such as quantum tomography to characterize the device and verify its underlying quantum mechanical structure, without performing the full Boson sampling task~\cite{deng2023gaussian}. However, just because the computational complexity class suggests that the problem is ``easy'' for a classical computer, still the question arises what the practical limits of this approach are. To that end, high-performance computing (HPC) is a very well established field which has great potential to assist in these quantum tomography tasks, provided the benefits of parallelisation can be reconciled with the constraints imposed by the quantum mechanical objects to be reconstructed. 

Quantum detector tomography~\cite{luis1999complete,fiurasek2001maximum,dariano2004quantum,lundeen2009tomography,feito2009measuring,coldenstrodt2009proposed} is a well-established technique for providing a consistent quantum mechanical characterisation of the measurement process. This approach to characterising experiments is particularly attractive, since it provides a model-free method to connect the underlying quantum mechanics of systems to the measurement results we observe. The aim of a tomography experiment is to reconstruct the set of Positive Operator Valued Measures (POVMs) $\{\mat{\pi}_n\}$ by mapping the detector response to a tomographically complete set of input states, {i.e.} the set of input states that span the full outcome space of the detector. The size of the problem is governed by the dimensionality of the Hilbert space $M$ occupied by the set of input states and the number of outcomes $N$. In order to be tomographically complete, the Hilbert space spanned by the input states is necessarily at least as large as the outcome space, {i.e.} $M\gtrsim N$. In general, the size of the set of POVMs $\{\mat{\pi}_n\}$ is then $M^2 \cdot N$; for non-phase-sensitive detectors this reduces to $M\cdot N$. Nevertheless, the challenge is thus to devise techniques to reconstruct POVMs covering ever larger system sizes, to enable state of the art quantum optics experiments~\cite{oripov2023a,deng2023gaussian}.

Up to now, almost all detector tomography experiments have described detectors with few outcomes ($N\lesssim10$) covering a relatively small Hilbert space ($M\lesssim100$)~\cite{lundeen2009tomography,feito2009measuring,brida2012quantum,humphreys2015tomography,schapeler2020quantum,endo2021quantum,cai2021quantum,fitzke2022time,santana2023extending_preprint}, where approaches such as semi-definite programming and maximum likelihood estimation could be readily applied with standard computing hardware. Alternatively, data pattern tomography can be used to characterise a relevant subset of the detector's outcome space~\cite{cooper2014local}. More recently, numerical approaches using convex optimisation solvers have been pursued, to interrogate larger system sizes~\cite{schapeler2021quantum}. This approach has been applied to high-performance computing hardware, as investigated by Liu et al.,~\cite{liu2023optimized} using simulated data. Their results suggested an upper limit of system size $M\cdot N$ of the order $10^5$, based on available computational resources. In the case of phase sensitive detectors, the size of the matrix required to map the Hilbert space dimension $M$ becomes $M^2$, significantly increasing the computational resource requirements. In this context, the largest tomographic reconstruction to date has been performed on a phase-sensitive photon counter, requiring the reconstruction of $1.8\cdot10^6$ elements~\cite{zhang2012mapping}. 

In this work, we perform experimental detector tomography of a high dynamic range detector up to $N=50$ occupied outcomes covering a Hilbert space of $M=1.2\cdot 10^6$ photons, surpassing the limit suggested by Liu et al.~\cite{liu2023optimized} as well as the size of the experimental reconstruction by Zhang et al.~\cite{zhang2012mapping}, by two orders of magnitude. To do so, we developed a convex minimisation solver optimised for operation on high-performance computing hardware. We apply this solver to the simulated data from Liu et al. (340 outcomes, Hilbert space up to $3\cdot10^4$ photons), and show two orders of magnitude saving in runtime, and four orders of magnitude saving in memory usage. Extending the simulated data further, we demonstrate reconstructions of system sizes $>10^{12}$ (e.g. $10^6$ outcomes, $10^6$ photons). Among many other applications, this allows for a quantum characterisation of state-of-the-art single-photon sensitive detector arrays~\cite{oripov2023a,morimoto2020megapixel} and Boson sampling machines~\cite{deng2023gaussian}, in minutes of computation time.

\section{Results}
\subsection{Tomographic approach}
\begin{figure}
    \centering
    \includegraphics[width=\textwidth]{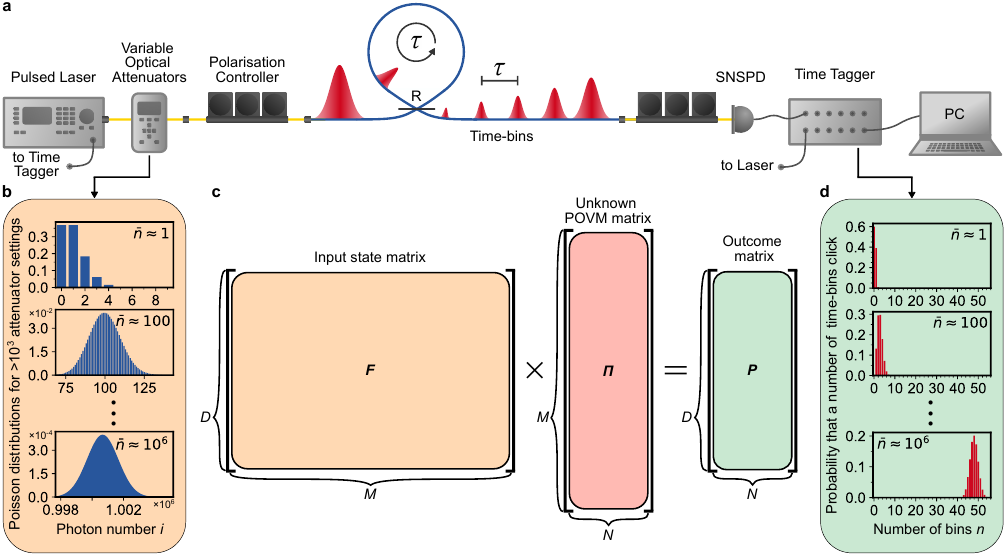}
    \caption{(a) Experimental setup to perform high dynamic range quantum detector tomography. The coherent states from a picosecond pulsed laser can be attenuated with variable optical attenuators, to control their mean photon numbers. Polarisation controllers before the fiber beam splitter loop (blue line) and the SNSPD are used to optimize the performance of the devices. The beam splitter loop creates sub-pulses with a temporal separation of $\tau=156~\mathrm{ns}$ and has an adaptable out-coupling $R$ and loop-efficiency $\eta_{\mathrm{loop}}$. A time tagger records raw time tags of the electrical output signal of the SNSPD. (b) Poisson distributions for mean photon numbers of $\bar{n}=1$, $\bar{n}=100$ and $\bar{n}=10^6$, which are set by the variable optical attenuator. (c) Schematic matrix representation of the matrices $\mat{F}_{D\times M}$ containing the coherent input states, $\mat{\Pi}_{M\times N}$ containing the (unknown) POVMs of the detector and $\mat{P}_{D\times N}$ containing the measured outcomes of the detector. (d) Probability distributions that a certain number of time-bins of the detector click, i.e., the outcomes of the detector for mean photon numbers of $\bar{n}=1$, $\bar{n}=100$ and $\bar{n}=10^6$.}
    \label{fig:approach}
\end{figure}

The principle of operation of detector tomography is depicted in Fig.~\ref{fig:approach}. Given a tomographically complete set of input states (beige), many measurements are performed, resulting in a set of outcome probabilities (green) related to the outcomes via the set of POVMs (red). Quantum mechanically, measurement outcomes $p_{\mat{\rho},n}$ on each state $\mat{\rho}$ via the POVM $\mat{\pi}_n$ are governed by the Born rule $p_{\mat{\rho},n}=\mathrm{Tr}\left[\mat{\rho} \mat{\pi}_n\right]$. Taking all measurements together, one can recast the Born rule as the matrix equation
\begin{equation}\label{eq:Matrix}
    \mat{P}=\mat{F}\mat{\Pi}
\end{equation}
where the matrix of outcome probabilities $\mat{P}_{D\times N}$ (green) are related to the set of input states $\mat{F}_{D\times M}$ (beige) via the set of POVMs $\mat{\Pi}_{M\times N}$ (red). The experimental setup used to generate and measure the set of input states is shown in Fig.~\ref{fig:approach}, with experimental details given in the Methods section.

In principle, one can invert Eq.~\eqref{eq:Matrix} to recover the unknown matrix of POVMs $\mat{\Pi}$.
In practice, however, the direct inversion of Eq.~\eqref{eq:Matrix} often yields nonphysical results, {i.e.} POVMs which are not positive semi-definite and/or unit trace. This is due to the presence of unavoidable noise in the experimental outcomes. 
To overcome this, one can recast Eq.~\eqref{eq:Matrix} into the constrained minimisation problem
\begin{subequations}
\begin{eqnarray}
    \min ||\mat{P}-\mat{F}\mat{\Pi}||_2 \label{eq:min_obj}\,,\\
    \textrm{subject to } \mat{\pi}_n\succeq0\,,~\sum_{n=0}^{N-1}\mat{\pi}_n=\mat{I} \label{eq:min_cons}\,,
\end{eqnarray}
\end{subequations}
which searches for the set of parameters in the POVM matrix $\mat{\Pi}$ which, when multiplied by the set of input states $\mat{F}$, most closely match the measured data $\mat{P}$. The two-norm, $||\mat A||_2=\sqrt{\sum_{ij} A_{i,j}^2}$, is used and $\mat{A}\succeq 0$ denotes the positive semi-definiteness constraint of the matrix $\mat{A}$.

The solution of the minimisation problem Eq.~\eqref{eq:min_obj} with the constraints Eq.~\eqref{eq:min_cons} for large POVM matrices is a numerically challenging convex optimization problem. Furthermore, the structure of the problem and the constraints render parallelisation nontrivial, since an efficient communication scheme between nodes is required.
The solution's computational effort and memory usage mainly depend on the number of free variables, {i.e.} the size of the matrix of POVMs $\mat{\Pi}$. The construction of $\mat{\Pi}$ from $\mat{\pi}_n$, and therefore the matrix dimensions, depend on whether or not the detector is sensitive to coherence present in the input states. In what follows, we consider phase insensitive detectors, the POVMs $\mat{\pi}_n$ of which can be expressed as diagonal matrices, with elements $\left(\pi_n\right)_{k,l}$ nonzero for $k=l$. For each $\mat{\pi}_n$, these nonzero elements comprise each row of the matrix of POVMs $\mat{\Pi}$, {i.e.} $\Pi_{i,n}=\left(\pi_{n}\right)_{k=i,l=i}.$ In this case, the size of the matrix and the number of entries in the matrix $\mat{\Pi}$ is thus $M\cdot N$, where $M-1$ denotes the maximal photon number in the Hilbert space with dimension $M$, and $N$ is the number of detector outcomes. Furthermore, in this case the constraints on $\mat{\pi}_n$ are such that the elements of $\mat{\Pi}$ are non-negative.
The constrained minimisation problem in the conventional form is thus
\begin{subequations}\label{eq:min}
\begin{eqnarray}
    \min_{\mat{\Pi} \in \mathbb{R}^{M \times N}}& ||\mat{P}-\mat{F} \mat{\Pi}||_2^2 \label{eq:min_obj2}\\
    \Pi_{i,n} \geq 0 \ \forall \ i\in \{0,...,M-1\},&\ n \in \{0,...,N-1\}\label{eq:min_ineq_cons2}\\
    \sum_{n=0}^{N-1} \Pi_{i,n} =1 \ & \forall \ i\in \{0,...,M-1\}\label{eq:min:eq_cons2}
\end{eqnarray}
\end{subequations}
for a given $\mat{P}\in \mathbb{R}^{D\times N}$ and given $\mat{F}\in \mathbb{R}^{D\times M}$. $D$ denotes the number of probe states. The memory needed for storing $\mat{\Pi}$, $\mat{F}$ and $\mat{P}$ is roughly
\begin{eqnarray}
    \mathrm{mem}_\mathrm{storage}&= 8\ \mathrm{byte} \cdot (MN+DM+DN).
\end{eqnarray}

The recent work by Liu et al.~\cite{liu2023optimized} has shown the practical solution to this problem for the order of $10^5$ free variables using the general minimisation solver MOSEK~\cite{mosek} via CVXPY~\cite{diamond2016cvxpy,agrawal2018rewriting}. The detector setup in their work requires $M$ and $D$ for a given number of outcomes $N$ to be approximately
\begin{eqnarray}
    M_\mathrm{Liu}&\approx 6.6\cdot N^{1.06} \label{eq:liu_M}\\
    D_\mathrm{Liu}&\approx 4.0\cdot N^{1.14} \label{eq:liu_D}.
\end{eqnarray}
Thus, the memory usage in the situation of Liu at al. for storing $\mat{\Pi}$, $\mat{F}$ and $\mat{P}$ is
\begin{eqnarray}
    \mathrm{mem}_\mathrm{storage}&\lesssim 2.6 \cdot 10^{-7} \cdot N^{2.2} \ \mathrm{GiB} \label{eq:mem_sdt_storage}.
\end{eqnarray}

Using our high-performance computing hardware~\cite{noctua2}, we investigate the practical main memory usage (see Appendix~\ref{app:software} for hardware and memory usage benchmark) of the solution approach of Liu et al.~\cite{liu2023optimized} for the standard detector tomography (SDT) problem and the simplified modified detector tomography (MDT) problem. Fitting the measured memory usage to the model $\mathrm{mem}=aN^b$ results in
\begin{eqnarray}
    \mathrm{mem}_\mathrm{Liu,SDT}&\approx 1.8 \cdot 10^{-5}\cdot N^{2.95} \ \mathrm{GiB}\label{eq:mem_liu_sdt_scaling} \\
    \mathrm{mem}_\mathrm{Liu,MDT}&\approx 2.0 \cdot 10^{-5}\cdot N^{2.93} \ \mathrm{GiB}~. 
    \label{eq:mem_liu_mdt_scaling}
\end{eqnarray}
This differs slightly from the scaling approximation given in Ref.~\cite{liu2023optimized}; on the one hand, we track the memory usage of the solver continually, and on the other, we use a wider range of total outcomes $N=11,21,...,201$.

Notwithstanding these slight differences, the prefactors in the memory usage scaling of the solver of Liu et al., Eq.~\eqref{eq:mem_liu_sdt_scaling}-\eqref{eq:mem_liu_mdt_scaling}, and the memory needed for the storage of matrices, Eq.~\eqref{eq:mem_sdt_storage}, are different by two orders of magnitude and the solver exhibits a scaling in $N$ that is significantly higher than for storage. This strongly suggests that a suitable numerical solver with a much lower memory footprint can be constructed to solve this problem more efficiently and scalable.

The goal of our approach is, on the one hand, a memory usage that is only a few times larger than the memory required for storing $\mat{\Pi}$, $\mat{F}$ and $\mat{P}$ and, on the other hand, the possibility not only to use multiple CPU-cores of a single compute node, but to scale to very large problems by using the distributed memory of many compute nodes.
For this purpose, we propose a two-stage variant of the two-metric projected Newton approach~\cite{bertsekas1982,landi2008a,schmidt2011optimization} for the problem in Eq.~\eqref{eq:min}. The algorithm is described in detail in Appendix~\ref{app:algorithms}.

\subsection{Detector under test} \label{sec:DUT}
We apply our solver to an experimental tomographic dataset obtained from a high dynamic range optical detector, with sensitivity from the single-photon level up to bright light~\cite{tiedau2019high}. The high dynamic range is achieved using a multiplexing scheme in which an incoming optical pulse is split into sub-pulses of exponentially decreasing pulse energies, incident on a superconducting single-photon detector (as shown schematically in Fig.~\ref{fig:approach}(a)). The number of outcomes of the detector is governed by the number of sub-pulses which result in a measurement event. Since this detector has no intrinsic phase sensitivity, the POVMs are fully described by matrices diagonal in the number basis, given by $\mat{\pi}_n=\sum_{k=0}^{M-1}\theta^{(n)}_k|k\rangle\langle k|$. The aim of the reconstruction is thus to find all $\theta_k^{(n)}$ for the different detector outcomes $n$, up to a maximum photon number $M-1$, corresponding to a Hilbert space of size $M$.

To carry out the reconstruction, we need a set of input states which span the Hilbert space to which the detector is sensitive. For this task, it is convenient to use the set of coherent states $|\alpha\rangle=\sum_{i=0}^\infty e^{-\frac{|\alpha|^2|}{2}}\frac{\alpha^i}{\sqrt{i!}}|i\rangle$. These states provide a photon number distribution governed by Poissonian statistics defined by their mean photon number $|\alpha|^2$. We require a set of $D$ states of different mean photon number $|\alpha_d|^2$, such that the elements of the input state matrix may be written as
\begin{equation}\label{eq:Fmatrix}
    F_{d,i}=\langle i|\alpha_d\rangle\!\langle\alpha_d|i\rangle=\frac{|\alpha_d|^{2i}}{i!}e^{-|\alpha_d|^{2}}
\end{equation}
for all photon numbers $i\in[0,M-1]$. We limit ourselves to the diagonal elements of the density matrix $|\alpha\rangle\langle\alpha|$ since our detector is insensitive to phase.

While in principle a single mean photon number is sufficient to cover any Hilbert space, since the coefficients are nonzero for all photon numbers $i$, in practice one requires sufficient statistics for all photon numbers for a reliable reconstruction. We choose mean photon numbers to scale quadratically, i.e., $|\alpha_d|^2\approx d^2$.

Previously, we have conducted detector tomography of 11 outcomes of this device up to a Hilbert space of size $M\approx5\times10^3$~\cite{schapeler2021quantum}, limited by standard computational methods using CVXPY~\cite{diamond2016cvxpy,agrawal2018rewriting}. Nevertheless, the logarithmic response of the detector means that the Hilbert space to which it is sensitive goes far beyond this limit. Furthermore, the detector design is such that a relatively simple model of the device can be constructed~\cite{tiedau2019high}, which allows the POVMs to be derived analytically. This enables us to test the accuracy of the computational reconstruction up to arbitrary size, which is essential to ensure correct reconstruction with our solver.

\begin{figure}
    \centering
    \includegraphics[width=\textwidth]{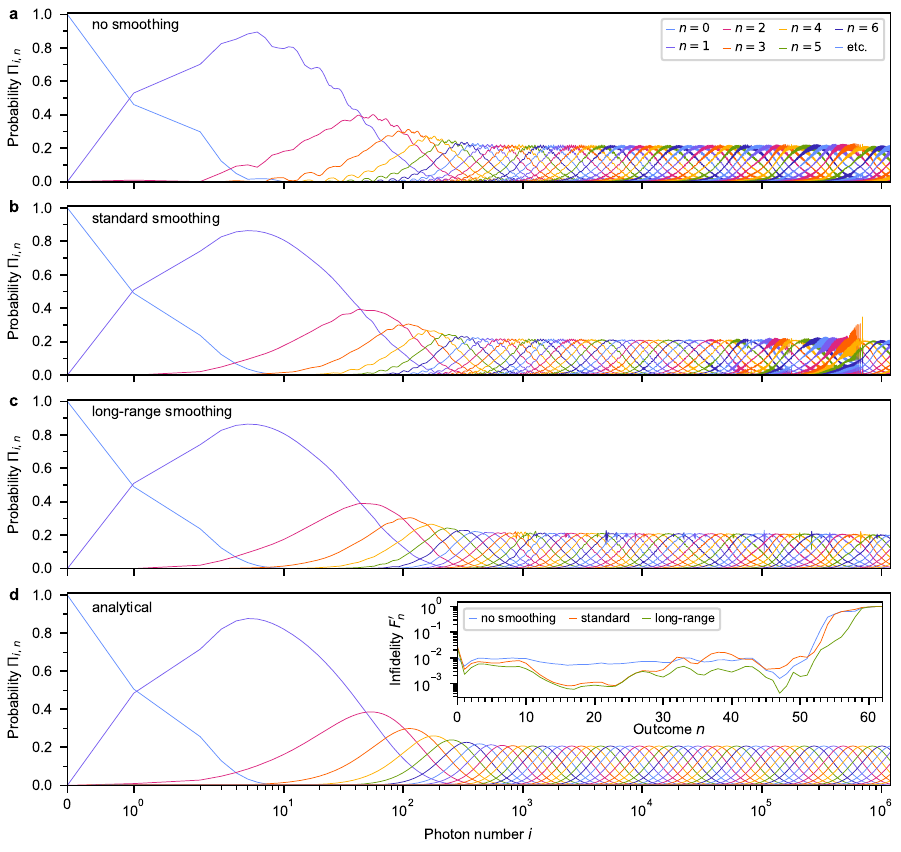}
    \caption{(a-c) Reconstructed POVMs from the experimentally measured data for $M=1210581$, $N=151$ outcomes, and $D=1076$. Due to the span of the input space, the first $\sim$50 outcomes are occupied and shown. (a) Does not include regularisation (smoothing) of the POVMs, (b) uses the standard nearest-neighbour smoothing with a regularisation parameter of $\gamma=10^{-5}$ and (c) in addition to the nearest-neighbour smoothing, utilises a novel long-range approach to the regularisation of the POVMs (see Methods, Regularisation and smoothing, for further detail). (d) Shows the analytical POVMs of the detector. The inset shows the infidelity Eq.~\eqref{eq:fidelities} between the three regularisation approaches and the analytical model for all occupied outcomes of the detector.}
    \label{fig:results}
\end{figure}

\subsection{POVM reconstruction}
We investigate the proposed reconstruction method with the experimental measurement data, i.e., with a Hilbert space cutoff of $M=1210581$, $N=151$ outcomes, of which the first $\sim$50 outcomes are tomographically covered by the $D=1076$ input states.
Figure~\ref{fig:results} shows three different tomographic reconstructions, each of which take different approaches to regularisation (see Methods, Regularisation and smoothing, for details on the regularisation routine). In Fig.~\ref{fig:results}(a) no regularisation is used, resulting in somewhat noisy response, particularly at low outcome numbers. In Fig.~\ref{fig:results}(b), a regularisation factor of $\gamma=10^{-5}$ is used, limited to smoothing between nearest-neighbour photon numbers, as discussed in Ref.~\cite{feito2009measuring}. This successfully smooths the high-frequency noise at low photon numbers, but is insufficient (and indeed increases noise) at high photon numbers. 

To mitigate this, we introduce a further step to the regularisation which encourages smoother, more physical results at higher photon numbers (see Methods, Regularisation and smoothing). The result of this is shown in Fig.~\ref{fig:results}(c).

To evaluate the solver in general, and each approach to regularisation in particular, we compare each reconstruction to the analytic model of the POVMs of the detector, shown in Fig.~\ref{fig:results}(d). In the inset of Fig.~\ref{fig:results}(d), we plot the infidelity (1-fidelity)
\begin{eqnarray}
    F_n^\prime=1-F_n=1-\frac{\mathrm{Tr}\left( \left(\sqrt{\mat{\pi}} \mat{\pi}_{n,\mathrm{theo}} \sqrt{\mat{\pi}} \right)^{1/2} \right) ^2}{\mathrm{Tr}(\mat{\pi}) \mathrm{Tr}(\mat{\pi}_{n,\mathrm{theo}})}\label{eq:fidelities}
\end{eqnarray}
between the reconstructions and the analytic model. Each case shows excellent agreement between the reconstructed and analytical POVMs, with fidelities exceeding 99\% for all occupied outcomes in all three cases. For the extended regularisation, this fidelity increases to an average of 99.69\% for all occupied outcomes up to $N=50$. This demonstrates the unparalleled accuracy of the reconstruction at unprecedented Hilbert space size. 

\begin{figure}
    \centering
    \includegraphics[width=\textwidth]{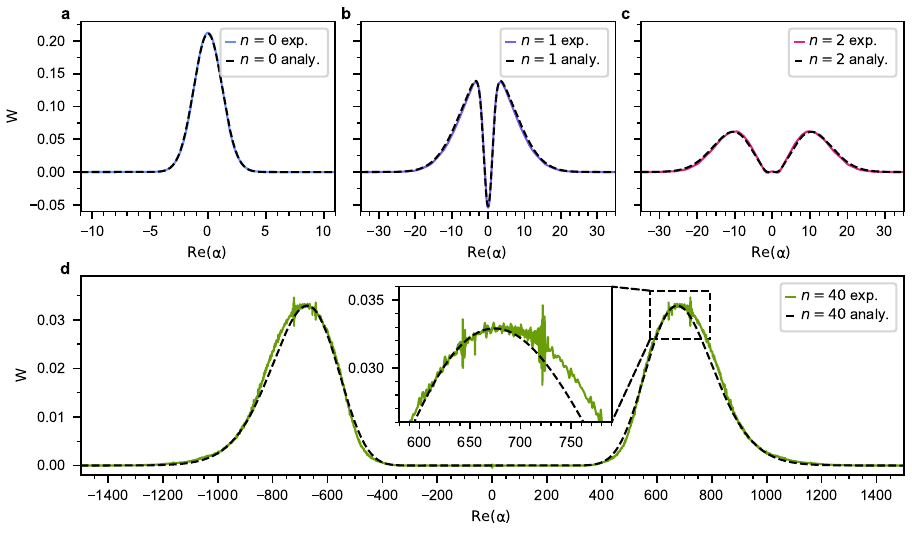}
    \caption{Wigner functions of the reconstructed POVMs using the long-range smoothing (shown in Fig.~\ref{fig:results}(c)) and the analytical POVMs (shown in Fig.~\ref{fig:results}(d)) for a subset of outcomes (a) $n=0$, (b) $n=1$, (c) $n=2$ and (d) $n=40$. Clear negativity in (b) shows the non-classical nature of the corresponding POVM $\mat{\pi}_{n=1}$. The inset shows that the general overlap of the Wigner functions based on the experimental (colored lines) and analytical (black dashed lines) POVMs is good, however, some noise appears in the Wigner functions for the experimental POVMs at larger outcomes.}
    \label{fig:wigner}
\end{figure}
From the POVMs of the detector we can reconstruct the Wigner functions corresponding to different detector outcomes, as shown in Fig.~\ref{fig:wigner}. These Wigner functions give additional insight about the detector. Negativity at the origin in Fig.~\ref{fig:wigner}(b), shows the non-classical nature of the corresponding operator $\mat{\pi}_{n=1}$. Our approach (see Methods, Computation of Wigner functions) for calculating the Wigner functions is stable even up to a Hilbert space dimension of $M=10^6$, as can be seen by the smooth nature of the Wigner functions of the analytical POVMs in Fig.~\ref{fig:wigner} (black dashed lines). Consequently, the noise in the Wigner function of the experimental POVM $\mat{\pi}_{n=40}$ can be explained by noise in the reconstructed POVM (which is visible in Fig.~\ref{fig:results}(c)). However, we see that the general overlap between Wigner functions of experimental and analytical POVMs is high, especially for small outcomes. 

\subsection{Application to other detector geometries}
\begin{figure}[ht!]
    \centering
    \includegraphics[width=\textwidth]{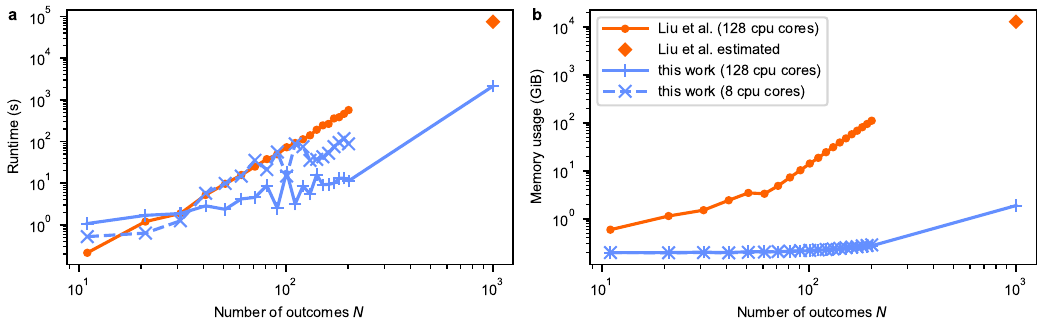}
    \caption{Runtimes (a) and memory usage (b) of the CVXPY-MOSEK-solver of Liu et al.~\cite{liu2023optimized} (orange dots) and the solver proposed in this work (blue crosses, 8 CPU cores; blue pluses, 128 CPU cores) for the detector setting of Liu et al. for different numbers of outcomes $N$. The orange diamond symbol show the estimated runtime and memory usage based on linear extrapolation for the solver of Liu et al. for $N=1000$.}
    \label{fig:liu}
\end{figure}

To compare with the current state of the art, we evaluate the proposed solution method and implementation for the detector geometry of Liu et al.~\cite{liu2023optimized}. They consider a spatially mutliplexed detector in the form of an on-chip network of beam splitters, whose output modes terminate in an SNSPD. This splits the input light equally onto many SNSPDs, thus gaining quasi-photon-number resolution. 
Their setting is characterized by the ratio of the maximal photon number in the Hilbert space and the number of outcomes, i.e., $M/N$ being of the order of only ten. Specifically, the relation in their case is given in Eq.~\eqref{eq:liu_M}. The inputs for the minimisation problem, i.e., the matrices $\mat F$ and $\mat P$ were generated with the implementation of Liu et al.~\cite{Liugithub}. In our analysis of their data, we use a single compute node with two AMD EPYC 7763 64-core CPUs. Further software and hardware details can be found in Appendix~\ref{app:software}.
The runtime and memory usage of the solver of Liu et al.~\cite{liu2023optimized} and the proposed method are shown in Fig.~\ref{fig:liu} for different $N$. While the CVXPY-MOSEK-based solver of Liu et al. has a lower runtime for small cases, i.e., $N\lesssim 30$, the solver proposed and implemented in this work is advantageous from intermediate sizes onwards. Performance of our proposed solver can be improved for small problems by using fewer CPU cores and, thus, threads which reduces the threading overhead. For $N=1000$, which corresponds to $N\cdot M=1.04\cdot 10^7$ free variables in $\mat \Pi$, the estimated runtime and memory usage for the CVXPY-MOSEK-based solver can be estimated to $\approx 75000$ seconds and $\approx 12700$ GiB while
the proposed solver requires $1450$ seconds and $1.6$ GiB. 

\subsection{Scalability for very large-scale problems}\label{sec:scalability_large}
The limiting factor for the problem size is the main memory available on the compute nodes. We choose to evaluate the so-called ``weak-scaling'' scenario, in which the portion of the problem that one node works on is chosen to be constant. In this case, the maximum problem size depends on the available memory per compute node, the number of nodes, and their communication (message passing interface (MPI) ranks - see Methods, Memory usage, for details). Using our high-performance computing hardware, the scalability of the approach up to $3.4\cdot 10^{12}$ free parameters has been demonstrated to be feasible, corresponding to about $27$ TB of storage for the POVM matrix $\mat{\Pi}$.

The question naturally arises as to how large a system can be reasonably reconstructed using this method. This cannot be answered in general since the iterative numerical solution depends on the number of iterations required to converge, which in turn depends on the specific problem to be solved as well as the experimental input data.
Thus, the scalability of the proposed solver and implementation is evaluated based on the scalability of the required performance-relevant operations ({e.g.} evaluation of the objective value, gradient, Hessian-matrix products, and others, as listed in Tab.~\ref{tab:ops} in the Methods) instead of the full reconstruction. 

Nevertheless, one can estimate the scaling of the solver based on heuristics.
Empirically for our two-stage iterative approach (see Methods, Algorithmic approach), the first stage requires about 10-15 Newton iterations, and the second stage requires about 30-200 Newton steps till convergence. Usually, between 20 and 50 conjugate-gradient iterations are needed per Newton iteration.
Thus, the runtimes measured for individual operations (presented in Fig.~\ref{fig:scaling_weak} in the Methods, Scalability considerations) for $N\cdot M$ up to $3\cdot 10^{12}$ can be multiplied by lower and upper estimates of iterations to give lower and upper estimates for the runtime of the reconstruction.

\begin{figure}[ht!]
    \centering
    \includegraphics{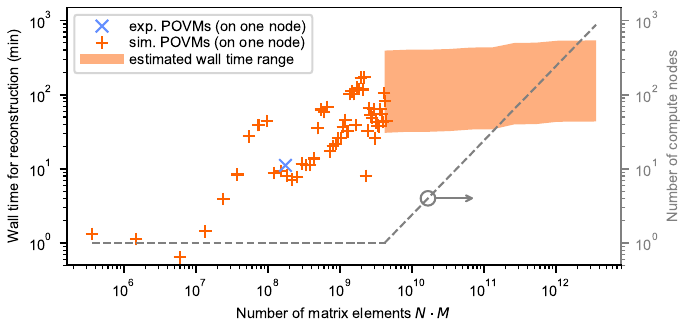}
    \caption{Scaling of the reconstruction wall time for a simulated detector with $N=151$ and $M=(D-1)^2$ up to one full compute node of the Noctua 2 cluster with 256 GiB of main memory ($N\cdot M\leq 4.4\cdot 10^9$). Beyond $N\cdot M\approx 4.4\cdot 10^9$ the wall time scaling for $N\approx 150$, $D=\sqrt{M}+1$ is estimated from the measurements of the scalability of the underlying operations like the gradient combined with the estimated numbers of operations described in Sec.~\ref{sec:scalability_large} assuming that the lowest possible number of compute nodes is used for the reconstruction, i.e., filling the main memory of the nodes. The blue cross represents the runtime for reconstructing the experimental POVMs using one compute node for comparison.}
    \label{fig:estimate_large}
\end{figure}

Given these results, in Figure~\ref{fig:estimate_large} we show estimated lower and upper limits of the runtime as applied to our detector geometry, up to a Hilbert space of $M\approx 3 \cdot 10^{12}$, number of outcomes $N\approx 150$ and requiring $D=\sqrt{M}$ input states to span the space. This is presented assuming that the number of iterations required for convergence remains similar to the experimentally realized case and that the number of compute nodes required to evaluate each system size is chosen according to the weak-scaling assumption (see Methods, Scalability considerations). The runtime for the reconstruction for the experimental setup of this work using one compute node of the Noctua 2 cluster~\cite{noctua2} is also shown for comparison.

\section{Discussion}\label{sec12}
Large-scale detectors for quantum light are becoming increasingly prevalent in quantum photonic technologies~\cite{oripov2023a,cheng2023a,eaton2023resolution}. The ability to accurately characterise these devices, in a consistent quantum mechanical framework is central to using them effectively. This approach has been used to debug photonic quantum computing platforms~\cite{deng2023gaussian}, and could be applied to large-scale single-photon sensitive imaging systems~\cite{oripov2023a,morimoto2020megapixel}. 
Developing computational tools to handle these large data sets, whilst preserving their quantum mechanical structure, is a vital task in the future proliferation of quantum photonic technologies.

We demonstrate the feasibility of applying high-performance computing to characterize a quantum detector that covers an extremely large outcome space, based on experimental data. The accuracy of this reconstruction is verified by comparison with an analytic model of the device, showing excellent agreement with fidelities above 99\% per outcome. Moreover, we are able to plot Wigner functions of these very large quantum objects by exploiting arbitrary-precision floating-point numbers, and this freely available code is generally applicable for diagonal density matrices~\cite{pqdts}.

The solver itself, which is also freely available~\cite{pqdts}, is designed for use with phase-insensitive detectors. The degree and type of regularisation can be adapted to the specific detector type. Extensions of the solver to cover phase sensitivity are also discussed in the Methods section. Furthermore, the solver can be modified to solve other problems of the form of Eq.~\eqref{eq:min}, provided the matrices fulfill the requirements (e.g. the banded structure of the $\mat{F}$-matrix) as outlined in the Methods section.

The main bottleneck of the solver is the available memory bandwidth. Thus, it might be worthwhile to consider and explore approximate computing techniques like reduced floating-point or fixed-point representations of numbers to further increase scale. In the other direction, using hardware with higher memory bandwidths like GPUs or FPGAs is a promising route.


\section{Methods}\label{Methods}

\subsection{Experiment}\label{Experiment}
The experimental setup is shown in Fig.~\ref{fig:approach}(a). We use a picosecond pulsed laser with a wavelength of $1556~\mathrm{nm}$ and a repetition rate of $25~\mathrm{kHz}$ to generate the coherent probe states. We can control the mean photon number per pulse by two variable optical attenuators that are placed after the laser. In order to ensure proper operation of the beam splitter loop (which is made from polarisation maintaining components) the polarisation is controlled by a manual fiber polarisation controller. Subsequently, the light pulses are coupled into the beam splitter loop with adaptable out-coupling $R$, loop-efficiency $\eta_{\mathrm{loop}}$ and temporal loop length $\tau$. The light is partially coupled out of the loop into the time-bins of the time-multiplexed detector. These sub-pulses have a separation of $\tau=156~\mathrm{ns}$, which needs to be larger than the dead time of the SNSPD. Before the detector another polarisation controller is used in order to optimize the polarisation dependent detection efficiency of the SNSPD. Given the repetition rate of the laser and the bin separation $\tau$ a maximum of 256 time-bins are allowed. 
We use a total of $D=1076$ different coherent probe states, whose mean photon numbers scale quadratically, in order to efficiently span the Hilbert space (with dimension $M\approx1.2\cdot10^6$) of the detector. We note that the experiment was limited by the pulse energy of the laser and not the detector itself, as in principle it is not possible to saturate this type of multiplexed detector. We record raw time tags with a time tagger for $5\cdot10^5$ trials of every input state.

The coherent probe state matrix $\mathbf{F}$ is constructed by expanding the coherent states $d\in[0,D-1]$ in the photon-number basis according to Eq.~\eqref{eq:Fmatrix} up to the Hilbert space dimension $i\in[0,M-1]$.

The outcome matrix $\mathbf{P}$ of the time-multiplexed detector is populated by counting the number of occupied time-bins per trial in a $5~\mathrm{ns}$ coincidence window and dividing by the total number of trials. This leads to the probabilities $P_{d,n}=\left.p_n\right\vert_d$ for different outcomes $n$ and input states $d$. We can truncate the number of time-bins at 150 (resulting in $N=151$ possible outcomes of the time-multiplexed detector), as subsequent time-bins are dominated by dark noise only.

As mentioned in Sec.~\ref{sec:DUT} it is possible to derive the POVMs of this high dynamic range detector analytically. In order to calculate the analytical POVMs, we first need to find the experimental parameters of the device. This can be done by fitting the bin-click probabilities, i.e., the probabilities that a certain bin $j$ fires for a given mean photon number. For a coherent state input to the detector, the bin-click probabilities are described by 
\begin{equation}\label{eq:pj_coh}
    p_j^{\mathrm{coh}}\left(d\right)=
    \Bigg\{\begin{array}{ll}
         1-
         \mathrm{exp}[-R\eta_{\mathrm{det}}|\alpha_d|^2]&j=1\\
         1-
         \mathrm{exp}\left[-(1-R)^2R^{-1}(R\eta_{\mathrm{loop}})^{j-1}|\alpha_d|^2\eta_{\mathrm{det}}\right]&j\geq2 
    \end{array}~,
\end{equation}
which we adapted from Ref.~\cite{tiedau2019high} to include the detector efficiency $\eta_{\mathrm{det}}$. We additionally neglect the dark-count probability, which is in the order of $5\times10^{-8}$ per time-bin in a $5~\mathrm{ns}$ coincidence window. We find the experimental parameters $R=0.91644(9)$, $\eta_{\mathrm{loop}}=0.90524(8)$ and $\eta_{\mathrm{det}}=0.528(1)$ by fitting Eq.~\eqref{eq:pj_coh} simultaneously for all coherent input states $|\alpha_d|^2$ with $d\in[0,D-1]$.

Given the experimental parameters of the high dynamic range detector, we can then calculate the bin-click probabilities for photon-number (Fock) state inputs
\begin{equation}\label{eq:pj_fock}
    p_j^{\mathrm{Fock}}\left(i\right)=
    \Bigg\{\begin{array}{ll}
         1-
         \left(1-R\eta_{\mathrm{det}}\right)^i&j=1\\
         1-
         \left[1-\left(1-R\right)^2R^{-1}
         \left(R\eta_\mathrm{loop}\right)^{j-1}\eta_{\mathrm{det}}\right]^i&j\geq2 
    \end{array}~,
\end{equation}
which is again adapted from Ref.~\cite{tiedau2019high} to include the detector efficiency $\eta_{\mathrm{det}}$. With these bin-click probabilities, it is possible to use a closed-form expression for the Poisson binomial distribution~\cite{fernandez2010closed} to calculate the POVMs of the detector (see Ref.~\cite{schapeler2021quantum} for more detail).

\subsection{High-performance computing approach}\label{hpc-approach}
\subsubsection{Algorithmic approach}\label{sec:algo}

Instead of relying on general minimisation solvers such as CVXPY~\cite{diamond2016cvxpy,agrawal2018rewriting} that are designed for handling arbitrary problems from large classes of minimisation problems, we propose a tailored algorithm that directly utilizes properties of the given problem. 

We have evaluated several different approaches that only require the objective function and constraints as well as their derivatives but no matrix factorizations or decompositions. The goal of this restriction was to set up an algorithm that is well suited to be parallelised not only for using multiple CPU cores in one compute node but multiple CPU-nodes of a large HPC-cluster in parallel. 
Thus, we have evaluated the augmented Lagrangian approach~\cite{powellmethod,Hestenes1969}, different variants of active-space methods as well as projection approaches~\cite{NoceWrig06}.

The minimisation problem at hand is characterized by a large number of equality ($M$) and inequality constraints ($N\cdot M$) and a relatively low computational complexity of the objective function due to the sparsity of $\mat{F}$. These characteristics directly impact the suitability of the classes of minimisation algorithms: For example, the large number of equality constraints cause a large number of additional augmentation and penalty terms in augmented Lagrangian-based approaches. 

Our implemented approach~\cite{pqdts} is based on the two-metric projected Newton method~\cite{bertsekas1982}. We developed a two-stage extension which showed significantly faster convergence. Both stages use a diagonally preconditioned conjugate gradient for the approximate solution of the linear system in Newton's method. As line search a backtracking approach with Armijo-like conditions~\cite{NoceWrig06} is used. 
The essential difference between the first stage and the second stage is that in the first stage a projection $\mathcal{P}_{\mathcal{S}^M}$ onto the $M$-dimensional unit simplex is used~\cite{Condat2016} whereas the second stage projects only onto the non-negativity constraints. Full details and algorithms are given in Appendix~\ref{app:algorithms}. In the following section we present a few important aspects of the implementation.

\subsubsection{Implementation}
\paragraph{Required operations}
The performance-relevant operations required for the solution approach described in Sec.~\ref{sec:algo} are listed in Table~\ref{tab:ops} together with their possible parallelism scope. The parallelism scope refers to the level or dimension along which the POVMs $\mat{\Pi}$ could be distributed to different processes or compute nodes without requiring communication or, if not possible, the additional required reduction is specified.
\begin{table}
\caption{The performance-relevant operations required for the approach described in section~\ref{sec:algo}. $\mat{c},\mat{d}\in \mathbb{R}^{M \times N}$, $\alpha \in \mathbb{R}$.}
    \centering
    \begin{tabular}{c|c|c}
        operation & equation & optimal parallelism \\ \hline
        objective value & $O(\mat{\Pi})=||\mat{P}-\mat{F}\mat{\Pi}||_2^2$ & columns of $\mat{\Pi}$ with $D\times N$-reduction \\
        gradient & $\partial_{\mat{\Pi}}O(\mat{\Pi})$ & columns of $\mat{\Pi}$ with $D\times N$-reduction \\
        Hessian products & $\mat{H}^{(k)} \mat{d}$ & columns of $\mat{\Pi}$ with $D\times N$-reduction \\
        diagonal of Hessian & $H^{(k)}_{i,j;i,j}=\frac{\partial^2 O(\Pi^{(k)})}{\partial \Pi_{i,j} \partial \Pi_{i,j}}$ &  columns of $\mat{\Pi}$ with $D\times N$-reduction  \\
        simplex-projection & $\mathcal{P}_{S^{M}}[\mat{d}]$ & rows of $\mat{d}$ \\
        element-wise operations & e.g. $\mat{d}+\mat{c}$,$\mat{c}\cdot \mat{d}$  & rows and/or columns of $\mat{d}$\\
        scalar multiplications & $\alpha\mat{d}$  & elements of $\mat{d}$\\
        2-norm & $\sqrt{\sum_{i,n}d_{i,n}^2}$  & rows/columns of $d_{i,j}$ with global reduction\\
        row-maxima & $m(i)=\mathrm{arg}\max_{n \in \{0,...,N-1\}}d_{i,n}$  & rows of $\mat{d}$\\
        row-sums & $s(i)=\mathrm{arg}\sum_{n=0}^{N-1}d_{i,n}$  & rows of $\mat{d}$\\
    \end{tabular}
    \label{tab:ops}
\end{table}

\begin{figure}
    \centering
    \includegraphics{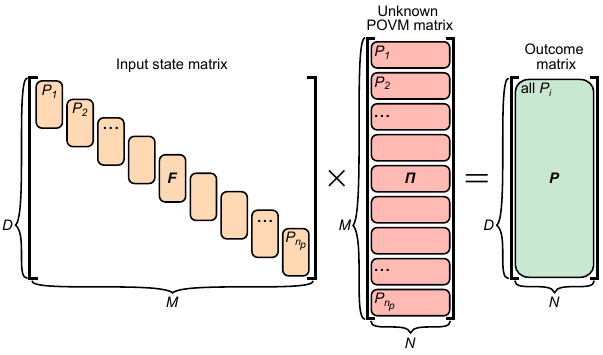}
    \caption{Schematic representation of the distribution of $\mat{F}$, $\mat{\Pi}$, and $\mat{P}$ to the processes $P_1$,...,$P_{n_p}$. Colored areas show distribution units in parallelisation.}
    \label{fig:PFX}
\end{figure}

\paragraph{Data distribution and parallelisation}
While all operations involving the objective function $O(\mat{\Pi})=||\mat{P}-\mat{F}\mat{\Pi}||_2^2$ have the columns of $\mat{\Pi}$ as a natural parallelism level due to the product $\mat{F} \mat{\Pi}$, all other required operations have at least the rows of $\mat{\Pi}$ as a natural parallelism level. Thankfully, for large $M$, the matrix $\mat{F}$ is sparse and banded, making it possible to also efficiently parallelise the operations that include the objective functions with respect to the rows of $\mat{\Pi}$. Thus, $\mat{\Pi}$ is distributed in blocks of rows to different processes. The matrices $\mat{P}$ and $\mat P-\mat{F}\mat{\Pi}$ are replicated on every compute node. Only the relevant non-zero blocks of $\mat{F}$ are stored on the corresponding processes. The distribution schema is schematically shown in Fig.~\ref{fig:PFX}.
Thus, the computation of $\mat F \mat \Pi$ is performed in two steps: The first step is the process-local computation of the contributions to the auxiliary matrix $\mat O=\mat F \mat \Pi$ with the rows of $\mat \Pi$ and the blocks of $\mat F$ that are present on the process. Secondly, the locally calculated contributions to $\mat{O}$ must be communicated to other processes. Due to the sparse banded structure of $\mat F$ for large $M$ and because only parts of $\mat O$ are required on every process to compute the objective function, gradient, and Hessian products, no all-to-all communication is required for this step. 
Instead, the communication pattern can be handled efficiently with a butterfly graph.
With the described distribution of blocks of rows of $\mat{\Pi}$, the proposed reconstruction can scale to very large problem sizes because the entire main memory available in a large HPC cluster can be used to compute a reconstruction.

We have implemented the distributed memory parallelisation with the message-passing interface (MPI)~\cite{mpi41}.
Additionally, we have parallelised the process-local operations with OpenMP~\cite{openmp} wherever possible. An emphasis was placed on avoiding false sharing of cache lines between the threads and an optimal usage of the available memory bandwidth. 

\subsubsection{Memory usage}
The limiting factor for the size of the reconstruction in terms of the problem parameters $N$, $M$, and $D$ that can be performed is the sum of the main memory available on the compute nodes.

The maximal possible Hilbert space dimension $M_\mathrm{max}$ possible for a reconstruction with our approach for a given number of outcomes $N$, number of probe states $D$, number of MPI-ranks $n_\mathrm{ranks}$ per compute node and number of compute nodes $n_\mathrm{nodes}$ and main memory per node $\mathrm{mem}_\mathrm{node}$ can be estimated with
\begin{eqnarray}
M_\mathrm{max}\approx \frac{n_\mathrm{nodes}}{6}\left(\frac{\mathrm{mem}_\mathrm{node}}{N\cdot 8 \ \mathrm{byte}}-2DN_\mathrm{ranks} \right).\label{eq:sclaing_N_mem}
\end{eqnarray}

\subsubsection{Scalability considerations}
Due to the dependence of the number of iterations during the minimisation on the input data, general statements on overall time-to-solution are limited. Nevertheless, the underlying operations can be analysed in detail.
We have chosen the quadratic scaling of probe states, i.e., $M=(D-1)^2$ corresponding to the experimental situation presented in this work. The matrices $\mat{P}$ and the POVMs $\mat{\Pi}$ were chosen randomly and densely for the scalability experiments. 

To analyse the scalability for very large reconstructions we chose a situation where the main memory of the available compute nodes is almost completely used and, thus, the limiting factor. Thus, the situation is related to the so-called weak-scaling case in parallelism where the portion of the problem that one compute node works on is kept constant when the number of compute nodes is increased.

For the compute nodes of Noctua 2~\cite{noctua2} with a main memory size of 256 GiB we have used $\mathrm{mem}_\mathrm{node}=200\ \mathrm{GB}$ to also account for buffer sizes of MPI transfers, OpenMP stack usage and other additional memory usages that are not covered in the approximation of Eq.~\eqref{eq:sclaing_N_mem}. Scalability was investigated for 8 MPI ranks per node, i.e., one per NUMA domain, and 16 threads per MPI rank.

\begin{figure}
    \centering
    \includegraphics[width=\textwidth]{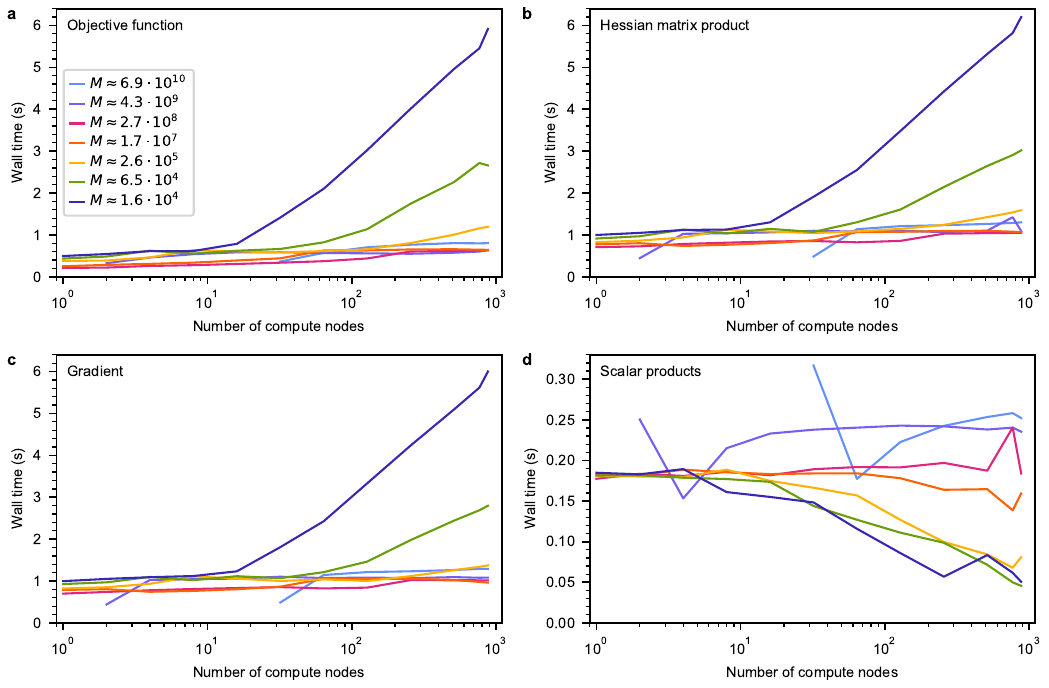}
    \caption{Weak-scaling behavior of the runtime for the evaluation of the (a) objective function, (b) the Hessian-matrix product, (c) gradient and (d) scalar products.}
    \label{fig:scaling_weak}
\end{figure}

The scaling behavior of underlying operations is shown in Fig.~\ref{fig:scaling_weak} for the evaluation of the objective function (a), the Hessian-matrix product (b), gradient (c) and scalar products (d), which all require global communication between the compute nodes. Runtime results for the gradient are nearly indistinguishable from the Hessian-matrix product. Element-wise operations, scalar multiplications, and projections are trivially parallel with the proposed parallelisation scheme. For example, a conjugate-gradient iteration requires one Hessian product, several element-wise operations, and several scalar multiplications. 

For larger photon number cutoffs $M\gtrsim 10^5$, the required runtime for the operations scales very favorably with the number of compute nodes. In contrast, for smaller $M$, the runtime increases visibly with the number of compute nodes. The underlying reason is that due to the choice of the weak-scaling scenario, in order to fill the main memory of the compute nodes in the case of small $M$, the number of outcomes $N$ has to be very high as determined from Eq.~\eqref{eq:sclaing_N_mem}. 

\paragraph{Regularisation and smoothing}
In addition to the least-squares objective function in Eq.~\eqref{eq:min_obj} our implementation also supports the next-neighbor regularisation term~\cite{lundeen2009tomography,feito2009measuring}
\begin{eqnarray}
    g(\mat \Pi)=\gamma \sum_{n=0}^{N-1}\sum_{i=0}^{M-2} (\Pi_{i,n}-\Pi_{i+1,n})^2 \label{eq:regul}
\end{eqnarray}
with the regularisation parameter $\gamma\ge0$.

While long-range generalizations of the regularisation term are, in principle, easily possible, they introduce an ambiguity due to the choice of the averaging and the range-dependence of the weighting. The minimisation problem for this work's experimental situation exhibits a flat valley around the minimum. Thus, we propose a different way of encouraging a more physical, i.e., less noisy reconstruction by first performing a reconstruction starting with some initial $\mat \Pi$. We use $\Pi_{i,n}^{(0)}=1/N$ as a starting point. The reconstruction problem is solved with the proposed two-stage algorithm and the result is smoothed by replacing each  $\Pi_{i,n}$ with its long-range average 
\begin{eqnarray}
    \tilde{\Pi}_{i,n}=\frac{1}{2N_\mathrm{smooth}(i)+1} \sum_{j=i-N_\mathrm{smooth}(i)}^{i+N_\mathrm{smooth}(i)}\Pi_{j,n}, \label{eq:smoothing}
\end{eqnarray}
where $\mat \Pi$ is the result of the minimisation and the smoothing distance $N_\mathrm{smooth}$ depends on the photon number $i$ so that the logarithmic photon-number scale in this experiment is accommodated.
The smoothed POVMs $\tilde {\mat{ \Pi}}$ then serve as the starting point for a second minimisation run where only the second stage is used. 

The exclusion of the lowest $\sim100$ photon numbers from the smoothing step improves the results because it prevents a severe disruption of the qualitative structure at low photon numbers. We have found that a smoothing distance of $N_\mathrm{smooth}=i/50$ yields a significant improvement over the POVMs without a smoothing step as shown in Fig.~\ref{fig:results}(c).

\paragraph{Possible improvements}
While our implementation assumes no a priori sparsity pattern of $\mat \Pi$, $\mat P$, and $\mat F$, a sparsity pattern forms in $\mat \Pi$ during the minimisation. However, our current implementation performs all operations, like the gradients or Hessian-matrix-products, in the full $N\cdot M$ space, which reduces the code complexity but is likely wasteful in terms of memory bandwidth and floating-point operations. Also, due to the experimental setting with $M\gg N$, the parallelisation and handling of matrices, especially $\mat F$, are optimized for this situation.

\subsubsection{Extension to phase-sensitive detectors}
The algorithm proposed in the previous section is tailored for the detector tomography reconstruction problem of a phase-insensitive detector given in Eq.~\eqref{eq:min}. For a phase-sensitive detector~\cite{feito2009measuring,zhang2012recursive,chen2022efficient}, the minimisation has the form
\begin{subequations}\label{eq:min_phase}
\begin{eqnarray}
    \min_{\mat{\pi}_n \in \mathbb{C}^{M \times M},\ n\in \{0,...,N-1\}}& ||\mat{P}-\mat{F} \mat{\Pi}||_2^2 \label{eq:min_phase_obj2}\\
    &\mat{\pi}_n=\mat{\pi}_n^\dagger &\ \forall \ n\in \{0,...,N-1\}\label{eq:min_phase_hermitian_cons2}\\
    &\mat{\pi}_n\succeq 0 &\ \forall \ n\in \{0,...,N-1\}\label{eq:min_phase_ineq_cons2}\\
    &\sum_{n=0}^{N-1} \mat{\pi}_n =\mathbb{1} \label{eq:min_phase_eq_cons2},
\end{eqnarray}
\end{subequations}
with $\mat{\Pi}=(\mat{\pi}_0,...,\mat{\pi}_{N-1})$. The two-metric projected Newton approach can be generalized to this situation by replacing the projection on non-negativity constraints with a projection of a hermitian matrix on its nearest positive semi-definite matrix. Such a projection can, in the conceptually most straight-forward way, be calculated as
\begin{eqnarray}
    \mathcal{P}_{\mathrm{semi-definite}}[\mat{A}]&=\mat{U}^\dagger \begin{pmatrix}
        \max(0,\lambda_1) & \\ & \max(0,\lambda_2) \\ & & \cdots
    \end{pmatrix}\mat{U},
\end{eqnarray}
where the unitary matrix $\mat{U}$ contains the eigenvectors of $\mat{A}$ as columns and $\lambda_i$ are the eigenvalues of $\mat{A}$. As this projection problem also arises in other areas, more efficient methods have been developed, especially for structured matrices~\cite{higham1988computing,francisco2017a}. 

\subsection{Computation of Wigner functions}\label{sec:wigner}
The computation of Wigner functions for high photon numbers is numerically challenging due to the high absolute values occurring during the calculation. For the high photon numbers in this work, the intermediate values can easily exceed the maximal values that are representable with conventional double-precision floating-point numbers. Thus, we have set up a variant~\cite{pqdts} by generalizing the implementation of the Wigner function available in QuantumOptics.jl~\cite{kramer2018quantumoptics} to use arbitrary-precision floating-point numbers via the BigFloat data type in Julia which uses the GNU MPFR library~\cite{fousse2007mpfr}. 

\backmatter
\bmhead{Data availability}
The raw data, as well as the relevant matrices and Wigner function data is openly available via Zenodo at \url{https://doi.org/10.5281/zenodo.10803758} and \url{https://doi.org/10.5281/zenodo.10810148}, respectively.

\bmhead{Code availability}
The reconstruction code ``Parallel Quantum Detector Tomography Solver'' (pqdts~\cite{pqdts}) and the code for calculating Wigner functions is openly available on GitHub via \url{https://github.com/pc2/pqdts} and can be cited via Zenodo \url{https://zenodo.org/doi/10.5281/zenodo.10853650}.

\bmhead{Acknowledgments}
The authors thank Lorenzo M. Procopio and Thomas Hummel for valuable discussion.
Funded by the European Union (ERC, QuESADILLA, 101042399). Views and opinions expressed are however those of the author(s) only and do not necessarily reflect those of the European Union or the European Research Council Executive Agency. Neither the European Union nor the granting authority can be held responsible for them. This work has received funding from the German Ministry of Education and Research within the PhoQuant project (grant number 13N16103).
The authors gratefully acknowledge the computing time provided to them on the high-performance computer Noctua 2 at the NHR Center PC2. These are funded by the Federal Ministry of Education and Research, and the state governments participate on the basis of the resolutions of the GWK for the national high-performance computing at universities (www.nhr-verein.de/unsere-partner).

\bmhead{Author contributions}
T.S. and T.B. conceived the idea. T.S. designed the experiment, performed the measurements and analysed the measurement data. T.S. and M.L. performed pretesting for the large-scale reconstruction. R.S. programmed the solver, performed all benchmarking and calculated the Wigner functions. T.S., R.S. and T.B. wrote the manuscript with inputs from all authors. C.P. and T.B. supervised the project.

\bmhead{Competing interest} 
The authors declare no competing interests.

\bmhead{Supplementary information}
Supplementary Information (see Appendix~\ref{app:software} and Appendix~\ref{app:algorithms}) is provided alongside the main document, where software and hardware details can be found, as well as the complete algorithmic description and additional information for the Wigner function computation. 

\begin{appendices}

\section{Software and hardware details}\label{app:software}
All numerical results have been obtained using compute nodes of the Noctua 2 high-performance computing cluster~\cite{noctua2}. Each node hosts two AMD EPYC 7763 CPUs with 64 CPU cores each and 256 GiB of DDR4-3200 main memory in an 8-channel-per-cpu configuration, giving a usable memory bandwidth of 370 GB/s in the STREAM Triad benchmark. The CPUs are configured with 4 NUMA domains per CPU, SMT is disabled, and turbo-mode is enabled.
The following software versions were used: Python 3.11.5, CVXPY 1.4.2, and MOSEK 10.1.24 for the solver of Liu et al. and GCC 13.2.0 with OpenMPI 4.1.4 for the proposed solver.

Main memory usage refers to the maximal resident set size of the solver process as measured by the Linux kernel as the high-water mark (HWM).

\section{Algorithms}\label{app:algorithms}
\subsection{Two-metric projected Newton method}\label{sec:solver_two_stage}
\paragraph{Bertsekas two-metric projected Newton method}
The two-metric projected Newton method for a general constrained minimisation problem
\begin{eqnarray}
    \min_{\vec x\in\Omega} f(\vec x), \label{eq:min_general_constrained}
\end{eqnarray}
where $\Omega$ is the space in which the constraints for $\vec x$ are fulfilled, is an iterative approach derived from the well-known Newton method for unconstrained minimisation~\cite{NoceWrig06}. 
For the unconstrained problem
\begin{eqnarray}
    \min_{\vec x\in\mathbb{R}^m} f(\vec x),\label{eq:min_general_unconstrained}
\end{eqnarray}
the unconstrained Newton method approximates the objective function $f(\vec x)$ in the $k$-th step by a quadratic model around the current iterate $\vec x^{(k)}$ as
\begin{eqnarray}
    Q^{(k)}(\vec x)&\approx f(\vec x^{(k)})+(\vec x-\vec x^{(k)})^T \nabla f(\vec x^{(k)})+\frac{1}{2}(\vec x-\vec x^{(k)})^T \mat{H}^{(k)} (\vec x-\vec x^{(k)}), \label{eq:newton_uncons_start}
\end{eqnarray}
where $\mat{H}^{(k)}$ is the Hessian matrix in the $k$-th step, i.e. $H_{ij}^{(k)}=\frac{\partial^2}{\partial x_i \partial x_j}f(\vec x^{(k)})$. 
The next iterate $\vec x^{(k+1)}$ is then the solution of the minimisation problem:
\begin{eqnarray}
    \vec x^{(k+1)}=\min_{\vec x \in \mathbb{R}^m} Q^{(k)}(\vec x). \label{eq:newton_uncons_min}
\end{eqnarray}
The quadratic unconstrained minimisation problem in Eq.~\eqref{eq:newton_uncons_min} can be rewritten as the well-known linear system for the minimisation step $\vec p^{(k)}=\vec x^{(k+1)}-\vec x^{(k)}$:
\begin{eqnarray}
    \mat{H}^{(k)}\vec p^{(k)}&=-\nabla f(\vec x^{(k)}), \label{eq:newton_linear}
\end{eqnarray}
which can be solved with various numerical methods. Usually, this schema is augmented by a line-search procedure where the next iterate $\vec x^{(k+1)}$ is determined from the step vector $\vec p^{(k)}$ by the solution of the line-search problem
\begin{eqnarray}
    \alpha^{(k)}=\mathrm{arg}\min_{\alpha \in \mathbb{R}} f(\vec x^{(k)}+\alpha \vec p^{(k)})\label{eq:linesearch}\\
    \vec x^{(k+1)}=\vec x^{(k)}+\alpha^{(k)} \vec p^{(k)} \label{eq:newton_uncons_end}.
\end{eqnarray} 
While this additional line search procedure only adds a typically negligible computational cost, the approach is much more numerically reliable. For an objective function with positive definite Hessian, the unconstrained Newton method converges quadratically fast near the solution~\cite{NoceWrig06}.

Analogous to the translation of the gradient descent method for unconstrained problems to the gradient projection method for constrained problems, the Newton method for unconstrained problems, Eq.~\eqref{eq:newton_uncons_start}-\eqref{eq:newton_uncons_end} can be translated to problems with constraints~\cite{NoceWrig06}. For this purpose, the unconstrained quadratic approximation in Eq.~\eqref{eq:newton_uncons_start} is written as a constrained problem that determines the next iterate $\vec x^{(k+1)}$ as
\begin{eqnarray}
    \vec x^{(k+1)}= \mathrm{arg}\min_{\vec x \in \Omega} Q^{(k)}(\vec x) \label{eq:newton_cons_start}.
\end{eqnarray}
the crucial difference is that the minimisation, in contrast to the unconstrained case in Eq.~\eqref{eq:newton_uncons_min}, is restricted to vectors $\vec x$ that fulfill the constraints. Consequently, the step direction $\vec p^{(k)}$ cannot be obtained by the solution of the linear system Eq.~\eqref{eq:newton_linear}, and instead, the constrained problem of Eq.~\eqref{eq:newton_cons_start} is typically significantly harder to solve. In the case of bound-constrained problems, i.e. $\Omega=\{\vec x \in \mathbb{R}^m: l \leq x_i \leq u \ \forall i\}$, one possibility to circumvent this issue is the idea of two-metric projected Newton approaches~\cite{bertsekas1982}. Equation~\eqref{eq:newton_cons_start} is rewritten as a projection
\begin{eqnarray}
    \vec x^{(k+1)}=\mathrm{arg} \min_{\vec x \in \mathbb{R}^m: l\leq x_i \leq u}\frac{1}{2} || \vec x-(\vec x^{(k)}-({\mat{H}^{(k)})}^{-1}\nabla f(\vec x^{(k)})) ||^2_{\mat{H}^{(k)}},
\end{eqnarray}
where the norm with respect to the metric defined by the positive-definite matrix $\mat A$ is defined as $||\vec y||_{\mat{A}}=\sqrt{\vec y^T\mat{A} \vec y}$.
Instead of solving this projection with respect to the metric induced by the Hessian matrix $\mat{H}$, the two-metric approach solves the projection problem
\begin{eqnarray}
    \vec x^{(k+1)}\approx \mathrm{arg} \min_{\vec x \in \mathbb{R}^m: l\leq x_i \leq u}\frac{1}{2} || \vec x-(\vec x^{(k)}-({\mat{D}^{(k)})}^{-1}\nabla f(\vec x^{(k)})) ||^2_{\mat{I}}\label{eq:2metric_unit_metric}
\end{eqnarray}
with respect to the unit metric, but the inverse of the Hessian matrix is replaced by the inverse of a suitable positive definite matrix $\mat D$. Practically, the projection with respect to the unit metric in Eq.~\eqref{eq:2metric_unit_metric} has the solution
\begin{eqnarray}
    \vec x^{(k+1)}\approx \mathcal{P}_{u,l}[\vec x^{(k)}-({\mat{D}^{(k)})}^{-1}\nabla f(\vec x^{(k)})],
\end{eqnarray}
where $\mathcal{P}_{u,l}(\vec x)$ is the projection of the vector $\vec x$ onto the constraint surface, i.e., 
\begin{eqnarray}
    (\mathcal{P}_{u,l}(\vec x))_i&=\begin{cases}
        l, \mathrm{if} \ x_i < l \\ 
        u, \mathrm{if} \ x_i > u \\
        x_i, \mathrm{otherwise}\end{cases}.
\end{eqnarray}
A line-search method is added to this approach as
\begin{eqnarray}
    \alpha^{(k)}=\mathrm{arg}\min_{\alpha \in \mathbb{R}} \mathcal{P}_{u,l}[\vec x^{(k)}-\alpha({\mat{D}^{(k)})}^{-1}\nabla f(\vec x^{(k)})]\\
    \vec x^{(k+1)}=\mathcal{P}_{u,l}[\vec x^{(k)}-\alpha^{(k)}({\mat{D}^{(k)})}^{-1}\nabla f(\vec x^{(k)})] \label{eq:2metric_alpha}
\end{eqnarray}
Bertsekas~\cite{bertsekas1982} has shown that for a problem with non-negativity constraints,
\begin{eqnarray}
    \min_{\vec x \in \mathbb{R}^m: x_i \geq 0 \ \forall i} f(\vec x)
\end{eqnarray}
a choice of the matrix $\mat{D}^{(k)}$ as 
\begin{eqnarray}
    ({D}^{(k)})_{ij}&=\begin{cases}
        0, \ \mathrm{if} \ i\neq j \ \mathrm{and} \ \mathrm{either}\ i \in I \ \mathrm{or} \ j \in I \\
        H_{ij}^{(k)}, \ \mathrm{otherwise}
    \end{cases}
\end{eqnarray}
with the index set $I$ defined as
\begin{eqnarray}
    I&=\{i: 0\leq x_i \leq \epsilon, \frac{\partial f(\vec x)}{\partial x_i}>0\} \label{eq:2metric_I}
\end{eqnarray}
and a small $\epsilon$, this approach leads to a globally convergent algorithm with a superlinear convergence rate under mild conditions.

For minimisation problems over a probability simplex, 
\begin{eqnarray}
    \min_{\vec x \in \mathbb{R}^m: x_i\geq 0 \ \forall i, \ \sum_{i=1}^m x_i=1 }f(\vec x),
\end{eqnarray}
Bertsekas~\cite{bertsekas1982} has proposed an approach that transforms the problem so that it can be handled with the two-metric projected Newton approach defined in Eq.~\eqref{eq:2metric_alpha}-\eqref{eq:2metric_I}. The transformation can be generalized straightforwardly to the tomography reconstruction problem in Eq.~\eqref{eq:min}, which has $M$ sum-constraints.
The approach only requires evaluations of the objective function, the gradient, and the approximate solution of the linear system
\begin{eqnarray}
    \mat{D}^{(k)} \vec p^{(k)}&=-\nabla f(\vec x^{(k)})
\end{eqnarray}
for $\vec p^{(k)}$, which can be implemented as an iterative preconditioned conjugate-gradient (CG) method where only matrix-product operations between $\mat{D}^{(k)}$ and vectors as well as vector-vector operations are required. Thus, the memory usage footprint is only a few times larger than the memory usage for the storage of the $\mat{P}$, $\mat{F}$, and $\mat{\Pi}$ given in Eq.~\eqref{eq:mem_sdt_storage}. The complete two-metric truncated Newton algorithm for the detector tomography problem Eq.~\eqref{eq:min} is given in Algo.~\ref{algo:2metric}.

\begin{algorithm}
\caption{Two-metric projected truncated Newton algorithm}
\begin{algorithmic}
\State $\Pi^{(0)}_{i,n} \gets \frac{1}{N}\ \forall i,n$ \Comment{Initialization that fulfills the constraints}
\State choose $0<\beta< 1$
\For{$k \gets 0,1,2,...$} \Comment{projected Newton iteration}
    \For{$i \gets 0,1,2,...,M$} \Comment{transformation step}
        \State $m(i) \gets \mathrm{arg}\max_{n\in\{0,...,N-1\}} \Pi^{(k)}_{i,n}$ \Comment{determine per-row maxima}
        \State $\tilde{\Pi}^{(k)}_{i,n} \gets \Pi^{(k)}_{i,n} \ \forall n \neq m(i)$
        \State $\tilde{\Pi}^{(k)}_{i,m(i)} \gets 1-\sum_{j\neq m(i)} \Pi^{(k)}_{i,j}$ \Comment{implicitly account for sum-constraint}
    \EndFor
    \State $\tilde{\mat{G}}^{(k)} \gets -\partial_{\tilde{\mat{\Pi}}^{(k)}} ||\mat{P}-\mat{F}\mat{\Pi}^{(k)}||_2^2$\Comment{gradient with respect to $\tilde{\mat{\Pi}}$}
    \State solve $\mat{D}^{(k)}\mat P^{(k)}=\tilde{\mat{G}}^{(k)}$ \Comment{with diagonally-preconditioned CG }
    \For{$l \gets 0,1,2,...$} \Comment{line search}
        \State $\alpha^{(k)} \gets \beta^l$
        \State ${\Pi}^{(k+1)}_{i,n} \gets \mathcal{P}_{0,\infty}[\tilde{\mat{\Pi}}^{(k)}+\alpha^{(k)} \mat{P}^{(k)}]_{i,n} \ \forall i,n\neq m(i)$ \Comment{update $\mat{\Pi}$}
        \State ${\Pi}^{(k+1)}_{i,m(i)} \gets 1-\sum_{j\neq m(i)} {\Pi}^{(k+1)}_{i,j}$
        \If{${\Pi}^{(k+1)}_{i,m(i)}\ge 0 \ \forall i$ \text{and} $\mat{{\Pi}}^{(k+1)}$ \text{fulfills Armijo-conditions~\cite{NoceWrig06}}}
            \State exit loop
        \EndIf
    \EndFor
    \If{\text{converged}}
        \State \text{exit loop}
    \EndIf
\EndFor
\end{algorithmic}
\label{algo:2metric}
\end{algorithm}

\begin{figure}[ht!]
    \centering
    \includegraphics{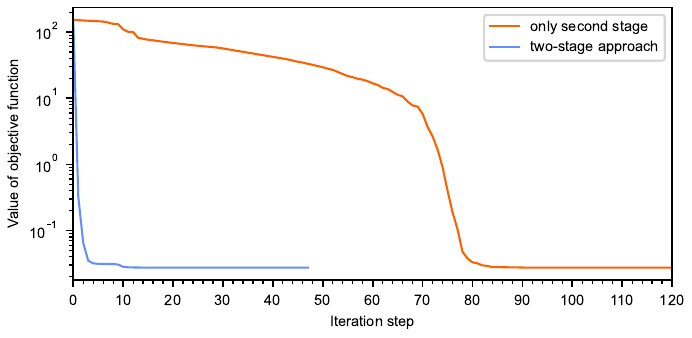}
    \caption{Comparison of the convergence of the proposed two-stage approach with the convergence when only using the second stage. No regularisation was applied, i.e., $\gamma=0$.}
    \label{fig:pqdts_exp_conv}
\end{figure}

\paragraph{Two-stage extension to the two-metric projected Newton approach}
While the algorithm described in the previous section has a favorable memory usage characteristic and can be parallelised efficiently, we have found that due to a large number of sum constraints, Eq.~\eqref{eq:min:eq_cons2}, the method converges slowly if it is not already rather close to the solution. The underlying reason is that in the line search in Algorithm~\ref{algo:2metric} the elements of $\mat{\Pi}^{(k+1)}$ corresponding to the row-wise maxima are obtained implicitly from the sum-constraint as 
\begin{eqnarray}
     {\Pi}^{(k+1)}_{i,m(i)}=1-\sum_{j\neq m(i)} {\Pi}^{(k+1)}_{i,j},
\end{eqnarray}
where $m(i)$ is the column-index of the maximum in the row $i$ of the matrix $\mat{\Pi}^{(k)}$. If $\mat{\Pi}^{(k)}$ is not close to the solution of the minimisation problem, there is likely a row in $\mat{\Pi}^{(k)}$ for which ${\Pi}^{(k)}_{i,m(i)}\ll 1$ and which, thus, forces the step length $\alpha^{(k)}$ to be small to fulfill the non-negativity constraint for ${\Pi}^{(k+1)}_{i,m(i)}$.
Thus, in the next section, we propose a two-stage approach that avoids this convergence slowdown. In the first stage, we use a modified variant of the two-metric projected Newton method of Bertsekas by replacing the transformation step for the sum constraints with a projection onto the probability simplex that also enforces the sum constraints. The projection $\mathcal{P}_{S}[\vec x]$ of a vector $\vec x \in \mathbb{R}^m$ onto the probability simplex $S=\{\vec y \in \mathbb{R}^m|y_i\ge 0, \sum_i y_i=1\}$ is defined as the point closest to $\vec x$ that is on $S$, i.e.,
\begin{eqnarray}
    \mathcal{P}_{S}[\vec x]=\mathrm{arg}\min_{\vec y  \in S} ||\vec x-\vec y||_2.
\end{eqnarray}
In practice, we use the algorithm proposed by Condat~\cite{Condat2016} that has a best-case complexity of $\mathcal{O}(m)$, a worst-case complexity of $\mathcal{O}(m^2)$, and additional memory requirement $\mathcal{O}(1)$.
For the purpose of the detector tomography reconstruction problem in Eq.~\eqref{eq:min}, an $M$-dimensional generalization $\mathcal{P}_{S^{M}}$ of projection on a probability simplex can be defined as
\begin{eqnarray}
    \mathcal{P}_{S^{M}}(\mat{\Pi})&=\begin{pmatrix}\mathcal{P}_{S}[(\Pi_{0,0},...,\Pi_{0,N-1})] \\ \vdots \\  \mathcal{P}_{S}[(\Pi_{M-1,0},...,\Pi_{M-1,N-1})] \end{pmatrix} \label{eq:Mp1_proj}
\end{eqnarray}
that projects the POVM $\mat{\Pi}$ on the constraints. The resulting algorithm is shown in Algo.~\ref{algo:2metric_mod}.
However, due to the presence of the $M$-dimensional projection $\mathcal{P}_{S^{M}}$, we have not yet been able to formally show global convergence for this modified variant and, thus, only use it as a first stage to accelerate convergence towards the solution. 
Once sufficiently close to the solution, we switch to the second stage with the proven globally convergent two-metric projected Newton approach in Algo.~\ref{algo:2metric}. Practical results that demonstrate the efficiency of this approach are shown in Fig.~\ref{fig:pqdts_exp_conv}
. The overall two-stage, two-metric projected Newton approach is shown in Algo.~\ref{algo:2metric_2stage}.
Thus, the two-stage algorithm is globally convergent and improves on the convergence issue of Algo.~\ref{algo:2metric}. The only required expensive operations are evaluating the objective function, gradient, and products of the modified Hessian $\mat{D}^{(k)}$ with vectors for the conjugate-gradient procedure. The constraints are enforced to numerical precision in both stages.
The memory usage of the algorithm is given as the sum of the memory needed to store the matrices $\mat F$, $\mat P$, $\mat \Pi^{(k)}$ plus the auxiliary matrices $\mat G^{(k)}$, $\mat O^{(k)}=\mat F \mat \Pi^{(k)}$, $\mat \Pi^{(k+1)}$, the index array $I$, and auxiliary matrices for the diagonally-preconditioned CG iteration. The overall memory requirement is
\begin{eqnarray}
    \mathrm{mem}_{\mathrm{2metric}}&= (2ND+6NM+MD+\mathcal{O}(M)+\mathcal{O}(N))\cdot 8\ \mathrm{byte}. \label{eq:2metric_mem}
\end{eqnarray}
For large $M$, a sparse storage of $\mat F$ instead of a dense representation is used to drastically reduce the term $MD$ in the memory estimation.
For comparison, in the case of the detector geometry of Liu et al., i.e., the dependence of $M$ and $D$ given by Eq.~\eqref{eq:liu_M}-\eqref{eq:liu_D}, and sufficiently large $N$, the memory estimate given in Eq.~\eqref{eq:2metric_mem} results in
\begin{eqnarray}
    \mathrm{mem}_{\mathrm{2metric},\mathrm{Liu}}&\lesssim 6.8 \cdot 10^{-7}\cdot N^{2.2}\ \mathrm{GiB} \label{eq:2metric_mem_liu}.
\end{eqnarray}

\begin{algorithm}
\caption{Two-metric projected truncated Newton algorithm}
\begin{algorithmic}
\State $\Pi^{(0)}_{i,n} \gets \frac{1}{N}\ \forall i,n$ \Comment{Initialization that fulfills the constraints}
\State choose $0<\beta< 1$
\For{$k \gets 0,1,2,...$} \Comment{projected Newton iteration}
    \State ${\mat{G}}^{(k)} \gets -\partial_{{\mat{\Pi}}^{(k)}} ||\mat{P}-\mat{F}\mat{\Pi}^{(k)}||_2^2$\Comment{gradient with respect to ${\mat{\Pi}}$}
    \State solve $\mat{D}^{(k)}\mat P^{(k)}={\mat{G}}^{(k)}$ \Comment{with diagonally-preconditioned CG}
    \For{$l \gets 0,1,2,...$} \Comment{line search}
        \State $\alpha^{(k)} \gets \beta^l$
        \State $\mat{\Pi}^{(k+1)} \gets \mathcal{P}_{S^{M+1}}[\mat{\Pi}^{(k)}+\alpha^{(k)} \mat{P}^{(k)}]$ \Comment{$M+1$-dimensional projection}
        \If{$\mat{{\Pi}}^{(k+1)}$ \text{fulfills Armijo-conditions}}
            \State exit loop
        \EndIf
    \EndFor
    \If{\text{converged}}
        \State \text{exit loop}
    \EndIf
\EndFor
\end{algorithmic}
\label{algo:2metric_mod}
\end{algorithm}

\begin{algorithm}
\caption{Two-stage, two-metric projected truncated Newton algorithm}
\begin{algorithmic}
\State $\Pi^{(0)}_{i,n} \gets \frac{1}{N}\ \forall i,n$ \Comment{Initialization that fulfills the constraints}
\State choose $0<\beta< 1$
\State minimize $\mat{\Pi}$ with Algo.~\ref{algo:2metric_mod} starting from $\mat{\Pi}^{(0)}$ \Comment{Stage 1}
\State minimize $\mat{\Pi}$ with Algo.~\ref{algo:2metric} starting from the result of the previous stage \Comment{Stage 2}
\end{algorithmic}
\label{algo:2metric_2stage}
\end{algorithm}

The convergence criterion for the line searches in both stages are the Armijo-like conditions where the smallest $m\in \mathbb{N}_0$ that fulfills
\begin{eqnarray}
    f(\mathcal{P}[\mat \Pi^{(k)}+\beta^m\mat P^{(k)}])\leq f(\mat \Pi^{(k)})+c\beta^m \frac{\partial}{\partial \alpha} f(\mathcal{P}[\mat \Pi^{(k)}+\alpha \mat P^{(k)}])|_{\alpha=0}
\end{eqnarray}
with the objective function $f$, $\beta=\frac{3}{4}$, and $c=\frac{1}{10}$ is used for the step length $\alpha^{(k)}=\beta^m$~\cite{bertsekas1982}.
For the second stage also the non-negativity conditions, $\Pi_{i,n}\geq 0$, have to be satisfied explicitly which is fulfilled implicitly for the first stage. 
For the convergence criterion for the transition from stage 1 to stage 2, the criterion
\begin{eqnarray}
|\partial_\alpha f(\mathcal{P}_{S^{M}}[\mat \Pi^{(k)}+\alpha \mat P^{(k)}])|\leq 10^{-4}
\end{eqnarray}
is chosen.
The convergence criterion for the second stage and, thus, for the overall algorithm is derived from the KKT-conditions~\cite{NoceWrig06} of the minimisation problem and defined as
\begin{eqnarray}
    \sqrt{\frac{1}{NM} \sum_{n=0}^{N-1} \sum_{i=0}^{M-1} (\Pi_{i,n}\cdot(\partial_{\Pi_{i,n}}f(\mat \Pi)+\lambda_{i}))^2}\leq \epsilon_\mathrm{KKT},
\end{eqnarray}
where $\lambda_i=\max(0,-\min_n{\partial_{\Pi_{i,n}}f(\mat \Pi)})$ is the estimate for the Lagrange multipliers of the sum-constraints.  

\paragraph{Convergence of the two-stage approach}
We investigate the convergence behavior of the proposed method with the experimental measurement data, i.e., with a Hilbert space cutoff of $M=1210581$, $N=151$ outcomes, and $D=1076$ probe states.

The effectiveness of the proposed two-stage approach can be seen by comparing the convergence speed when only using the second stage, i.e., Bertseka's two-metric projected Newton method in contrast to the two-stage approach depicted in Fig.~\ref{fig:pqdts_exp_conv}. While the convergence close to the minimum is very similar by construction, replacing the first iterations with the modified variant drastically accelerates convergence when far away from the minimum.

\subsection{Wigner function}
Computations with arbitrary-precision floating-point numbers required for the computation of the Wigner function in Sec.~\ref{sec:wigner} are much more expensive than double-precision floating-point numbers because no direct hardware implementation is available. We have optimized the implementation of the Wigner function for phase-insensitive detectors, i.e., diagonal density matrices, and have implemented a trivial parallelisation via MPI to scale the computation of the Wigner function to many CPU cores/compute nodes~\cite{pqdts}.

The arbitrary-precision floating-point numbers in MPFR are represented as the product of a sign $s\in \{-1,1\}$, a 64-bit exponent $e$, and an arbitrary-sized fixed-point mantissa $m$ as
\begin{equation}
    s\cdot 2^{e}\cdot m.
\end{equation}
The important ingredient is the very large range $\approx 10^{2^{62}}\approx 10^{10^{18}}$ of numbers that can be represented compared to the range of 64-bit floating-point numbers with a range of $\approx 10^{307}$. The bit size of the mantissa can be varied. For the Wigner functions in this work, mantissa sizes of about 60-70 bits are sufficient in the sense that larger mantissa sizes yield binary-identical Wigner functions when the final result is converted to double-precision floating-point numbers.

\end{appendices}

\bibliography{references}


\begin{thebibliography}{47}
\ifx \bisbn   \undefined \def \bisbn  #1{ISBN #1}\fi
\ifx \binits  \undefined \def \binits#1{#1}\fi
\ifx \bauthor  \undefined \def \bauthor#1{#1}\fi
\ifx \batitle  \undefined \def \batitle#1{#1}\fi
\ifx \bjtitle  \undefined \def \bjtitle#1{#1}\fi
\ifx \bvolume  \undefined \def \bvolume#1{\textbf{#1}}\fi
\ifx \byear  \undefined \def \byear#1{#1}\fi
\ifx \bissue  \undefined \def \bissue#1{#1}\fi
\ifx \bfpage  \undefined \def \bfpage#1{#1}\fi
\ifx \blpage  \undefined \def \blpage #1{#1}\fi
\ifx \burl  \undefined \def \burl#1{\textsf{#1}}\fi
\ifx \doiurl  \undefined \def \doiurl#1{\url{https://doi.org/#1}}\fi
\ifx \betal  \undefined \def \betal{\textit{et al.}}\fi
\ifx \binstitute  \undefined \def \binstitute#1{#1}\fi
\ifx \binstitutionaled  \undefined \def \binstitutionaled#1{#1}\fi
\ifx \bctitle  \undefined \def \bctitle#1{#1}\fi
\ifx \beditor  \undefined \def \beditor#1{#1}\fi
\ifx \bpublisher  \undefined \def \bpublisher#1{#1}\fi
\ifx \bbtitle  \undefined \def \bbtitle#1{#1}\fi
\ifx \bedition  \undefined \def \bedition#1{#1}\fi
\ifx \bseriesno  \undefined \def \bseriesno#1{#1}\fi
\ifx \blocation  \undefined \def \blocation#1{#1}\fi
\ifx \bsertitle  \undefined \def \bsertitle#1{#1}\fi
\ifx \bsnm \undefined \def \bsnm#1{#1}\fi
\ifx \bsuffix \undefined \def \bsuffix#1{#1}\fi
\ifx \bparticle \undefined \def \bparticle#1{#1}\fi
\ifx \barticle \undefined \def \barticle#1{#1}\fi
\bibcommenthead
\ifx \bconfdate \undefined \def \bconfdate #1{#1}\fi
\ifx \botherref \undefined \def \botherref #1{#1}\fi
\ifx \url \undefined \def \url#1{\textsf{#1}}\fi
\ifx \bchapter \undefined \def \bchapter#1{#1}\fi
\ifx \bbook \undefined \def \bbook#1{#1}\fi
\ifx \bcomment \undefined \def \bcomment#1{#1}\fi
\ifx \oauthor \undefined \def \oauthor#1{#1}\fi
\ifx \citeauthoryear \undefined \def \citeauthoryear#1{#1}\fi
\ifx \endbibitem  \undefined \def \endbibitem {}\fi
\ifx \bconflocation  \undefined \def \bconflocation#1{#1}\fi
\ifx \arxivurl  \undefined \def \arxivurl#1{\textsf{#1}}\fi
\csname PreBibitemsHook\endcsname

\bibitem[\protect\citeauthoryear{Aspuru-Guzik and Walther}{2012}]{aspuru2012photonic}
\begin{barticle}
\bauthor{\bsnm{Aspuru-Guzik}, \binits{A.}},
\bauthor{\bsnm{Walther}, \binits{P.}}:
\batitle{{Photonic quantum simulators}}.
\bjtitle{Nature Physics}
\bvolume{8}(\bissue{4}),
\bfpage{285}--\blpage{291}
(\byear{2012})
\doiurl{10.1038/nphys2253}
\end{barticle}
\endbibitem

\bibitem[\protect\citeauthoryear{Aaronson and Arkhipov}{2011}]{aaronson2011the}
\begin{bchapter}
\bauthor{\bsnm{Aaronson}, \binits{S.}},
\bauthor{\bsnm{Arkhipov}, \binits{A.}}:
\bctitle{{The computational complexity of linear optics}}.
In: \bbtitle{Proceedings of the Forty-third Annual ACM Symposium on Theory of Computing}.
\bsertitle{STOC '11},
pp. \bfpage{333}--\blpage{342}.
\bpublisher{ACM},
\blocation{New York, NY, USA}
(\byear{2011}).
\doiurl{10.1145/1993636.1993682}
\end{bchapter}
\endbibitem

\bibitem[\protect\citeauthoryear{Deng et~al.}{2023}]{deng2023gaussian}
\begin{barticle}
\bauthor{\bsnm{Deng}, \binits{Y.-H.}},
\bauthor{\bsnm{Gu}, \binits{Y.-C.}},
\bauthor{\bsnm{Liu}, \binits{H.-L.}},
\bauthor{\bsnm{Gong}, \binits{S.-Q.}},
\bauthor{\bsnm{Su}, \binits{H.}},
\bauthor{\bsnm{Zhang}, \binits{Z.-J.}},
\bauthor{\bsnm{Tang}, \binits{H.-Y.}},
\bauthor{\bsnm{Jia}, \binits{M.-H.}},
\bauthor{\bsnm{Xu}, \binits{J.-M.}},
\bauthor{\bsnm{Chen}, \binits{M.-C.}},
\bauthor{\bsnm{Qin}, \binits{J.}},
\bauthor{\bsnm{Peng}, \binits{L.-C.}},
\bauthor{\bsnm{Yan}, \binits{J.}},
\bauthor{\bsnm{Hu}, \binits{Y.}},
\bauthor{\bsnm{Huang}, \binits{J.}},
\bauthor{\bsnm{Li}, \binits{H.}},
\bauthor{\bsnm{Li}, \binits{Y.}},
\bauthor{\bsnm{Chen}, \binits{Y.}},
\bauthor{\bsnm{Jiang}, \binits{X.}},
\bauthor{\bsnm{Gan}, \binits{L.}},
\bauthor{\bsnm{Yang}, \binits{G.}},
\bauthor{\bsnm{You}, \binits{L.}},
\bauthor{\bsnm{Li}, \binits{L.}},
\bauthor{\bsnm{Zhong}, \binits{H.-S.}},
\bauthor{\bsnm{Wang}, \binits{H.}},
\bauthor{\bsnm{Liu}, \binits{N.-L.}},
\bauthor{\bsnm{Renema}, \binits{J.J.}},
\bauthor{\bsnm{Lu}, \binits{C.-Y.}},
\bauthor{\bsnm{Pan}, \binits{J.-W.}}:
\batitle{{Gaussian Boson Sampling with Pseudo-Photon-Number-Resolving Detectors and Quantum Computational Advantage}}.
\bjtitle{Physical Review Letters}
\bvolume{131}(\bissue{15}),
\bfpage{150601}
(\byear{2023})
\doiurl{10.1103/PhysRevLett.131.150601}
\end{barticle}
\endbibitem

\bibitem[\protect\citeauthoryear{Luis and S{\'{a}}nchez-Soto}{1999}]{luis1999complete}
\begin{barticle}
\bauthor{\bsnm{Luis}, \binits{A.}},
\bauthor{\bsnm{S{\'{a}}nchez-Soto}, \binits{L.L.}}:
\batitle{{Complete Characterization of Arbitrary Quantum Measurement Processes}}.
\bjtitle{Physical Review Letters}
\bvolume{83}(\bissue{18}),
\bfpage{3573}--\blpage{3576}
(\byear{1999})
\doiurl{10.1103/PhysRevLett.83.3573}
\end{barticle}
\endbibitem

\bibitem[\protect\citeauthoryear{Fiur{\'{a}}{\v{s}}ek}{2001}]{fiurasek2001maximum}
\begin{barticle}
\bauthor{\bsnm{Fiur{\'{a}}{\v{s}}ek}, \binits{J.}}:
\batitle{{Maximum-likelihood estimation of quantum measurement}}.
\bjtitle{Physical Review A}
\bvolume{64}(\bissue{2}),
\bfpage{24102}
(\byear{2001})
\doiurl{10.1103/PhysRevA.64.024102}
\end{barticle}
\endbibitem

\bibitem[\protect\citeauthoryear{D'Ariano et~al.}{2004}]{dariano2004quantum}
\begin{barticle}
\bauthor{\bsnm{D'Ariano}, \binits{G.M.}},
\bauthor{\bsnm{Maccone}, \binits{L.}},
\bauthor{\bsnm{Presti}, \binits{P.L.}}:
\batitle{{Quantum Calibration of Measurement Instrumentation}}.
\bjtitle{Physical Review Letters}
\bvolume{93}(\bissue{25}),
\bfpage{250407}
(\byear{2004})
\doiurl{10.1103/PhysRevLett.93.250407}
\end{barticle}
\endbibitem

\bibitem[\protect\citeauthoryear{Lundeen et~al.}{2009}]{lundeen2009tomography}
\begin{barticle}
\bauthor{\bsnm{Lundeen}, \binits{J.S.}},
\bauthor{\bsnm{Feito}, \binits{A.}},
\bauthor{\bsnm{Coldenstrodt-Ronge}, \binits{H.}},
\bauthor{\bsnm{Pregnell}, \binits{K.L.}},
\bauthor{\bsnm{Silberhorn}, \binits{C.}},
\bauthor{\bsnm{Ralph}, \binits{T.C.}},
\bauthor{\bsnm{Eisert}, \binits{J.}},
\bauthor{\bsnm{Plenio}, \binits{M.B.}},
\bauthor{\bsnm{Walmsley}, \binits{I.A.}}:
\batitle{{Tomography of quantum detectors}}.
\bjtitle{Nature Physics}
\bvolume{5}(\bissue{1}),
\bfpage{27}--\blpage{30}
(\byear{2009})
\doiurl{10.1038/nphys1133}
\end{barticle}
\endbibitem

\bibitem[\protect\citeauthoryear{Feito et~al.}{2009}]{feito2009measuring}
\begin{barticle}
\bauthor{\bsnm{Feito}, \binits{A.}},
\bauthor{\bsnm{Lundeen}, \binits{J.S.}},
\bauthor{\bsnm{Coldenstrodt-Ronge}, \binits{H.}},
\bauthor{\bsnm{Eisert}, \binits{J.}},
\bauthor{\bsnm{Plenio}, \binits{M.B.}},
\bauthor{\bsnm{Walmsley}, \binits{I.A.}}:
\batitle{{Measuring measurement: theory and practice}}.
\bjtitle{New Journal of Physics}
\bvolume{11}(\bissue{9}),
\bfpage{93038}
(\byear{2009})
\doiurl{10.1088/1367-2630/11/9/093038}
\end{barticle}
\endbibitem

\bibitem[\protect\citeauthoryear{Coldenstrodt-Ronge et~al.}{2009}]{coldenstrodt2009proposed}
\begin{barticle}
\bauthor{\bsnm{Coldenstrodt-Ronge}, \binits{H.B.}},
\bauthor{\bsnm{Lundeen}, \binits{J.S.}},
\bauthor{\bsnm{Pregnell}, \binits{K.L.}},
\bauthor{\bsnm{Feito}, \binits{A.}},
\bauthor{\bsnm{Smith}, \binits{B.J.}},
\bauthor{\bsnm{Mauerer}, \binits{W.}},
\bauthor{\bsnm{Silberhorn}, \binits{C.}},
\bauthor{\bsnm{Eisert}, \binits{J.}},
\bauthor{\bsnm{Plenio}, \binits{M.B.}},
\bauthor{\bsnm{Walmsley}, \binits{I.A.}}:
\batitle{{A proposed testbed for detector tomography}}.
\bjtitle{Journal of Modern Optics}
\bvolume{56}(\bissue{2-3}),
\bfpage{432}--\blpage{441}
(\byear{2009})
\doiurl{10.1080/09500340802304929}
\end{barticle}
\endbibitem

\bibitem[\protect\citeauthoryear{Oripov et~al.}{2023}]{oripov2023a}
\begin{barticle}
\bauthor{\bsnm{Oripov}, \binits{B.G.}},
\bauthor{\bsnm{Rampini}, \binits{D.S.}},
\bauthor{\bsnm{Allmaras}, \binits{J.}},
\bauthor{\bsnm{Shaw}, \binits{M.D.}},
\bauthor{\bsnm{Nam}, \binits{S.W.}},
\bauthor{\bsnm{Korzh}, \binits{B.}},
\bauthor{\bsnm{McCaughan}, \binits{A.N.}}:
\batitle{{A superconducting nanowire single-photon camera with 400,000 pixels}}.
\bjtitle{Nature}
\bvolume{622}(\bissue{7984}),
\bfpage{730}--\blpage{734}
(\byear{2023})
\doiurl{10.1038/s41586-023-06550-2}
\end{barticle}
\endbibitem

\bibitem[\protect\citeauthoryear{Brida et~al.}{2012}]{brida2012quantum}
\begin{barticle}
\bauthor{\bsnm{Brida}, \binits{G.}},
\bauthor{\bsnm{Ciavarella}, \binits{L.}},
\bauthor{\bsnm{Degiovanni}, \binits{I.P.}},
\bauthor{\bsnm{Genovese}, \binits{M.}},
\bauthor{\bsnm{Lolli}, \binits{L.}},
\bauthor{\bsnm{Mingolla}, \binits{M.G.}},
\bauthor{\bsnm{Piacentini}, \binits{F.}},
\bauthor{\bsnm{Rajteri}, \binits{M.}},
\bauthor{\bsnm{Taralli}, \binits{E.}},
\bauthor{\bsnm{Paris}, \binits{M.G.A.}}:
\batitle{{Quantum characterization of superconducting photon counters}}.
\bjtitle{New Journal of Physics}
\bvolume{14}(\bissue{8}),
\bfpage{85001}
(\byear{2012})
\doiurl{10.1088/1367-2630/14/8/085001}
\end{barticle}
\endbibitem

\bibitem[\protect\citeauthoryear{Humphreys et~al.}{2015}]{humphreys2015tomography}
\begin{barticle}
\bauthor{\bsnm{Humphreys}, \binits{P.C.}},
\bauthor{\bsnm{Metcalf}, \binits{B.J.}},
\bauthor{\bsnm{Gerrits}, \binits{T.}},
\bauthor{\bsnm{Hiemstra}, \binits{T.}},
\bauthor{\bsnm{Lita}, \binits{A.E.}},
\bauthor{\bsnm{Nunn}, \binits{J.}},
\bauthor{\bsnm{Nam}, \binits{S.W.}},
\bauthor{\bsnm{Datta}, \binits{A.}},
\bauthor{\bsnm{Kolthammer}, \binits{W.S.}},
\bauthor{\bsnm{Walmsley}, \binits{I.A.}}:
\batitle{{Tomography of photon-number resolving continuous-output detectors}}.
\bjtitle{New Journal of Physics}
\bvolume{17}(\bissue{10}),
\bfpage{103044}
(\byear{2015})
\doiurl{10.1088/1367-2630/17/10/103044}
\end{barticle}
\endbibitem

\bibitem[\protect\citeauthoryear{Schapeler et~al.}{2020}]{schapeler2020quantum}
\begin{barticle}
\bauthor{\bsnm{Schapeler}, \binits{T.}},
\bauthor{\bsnm{{Philipp H{\"{o}}pker}}, \binits{J.}},
\bauthor{\bsnm{Bartley}, \binits{T.J.}}:
\batitle{{Quantum detector tomography of a 2$\times$2 multi-pixel array of superconducting nanowire single photon detectors}}.
\bjtitle{Optics Express}
\bvolume{28}(\bissue{22}),
\bfpage{33035}--\blpage{33043}
(\byear{2020})
\doiurl{10.1364/OE.404285}
\end{barticle}
\endbibitem

\bibitem[\protect\citeauthoryear{Endo et~al.}{2021}]{endo2021quantum}
\begin{barticle}
\bauthor{\bsnm{Endo}, \binits{M.}},
\bauthor{\bsnm{Sonoyama}, \binits{T.}},
\bauthor{\bsnm{Matsuyama}, \binits{M.}},
\bauthor{\bsnm{Okamoto}, \binits{F.}},
\bauthor{\bsnm{Miki}, \binits{S.}},
\bauthor{\bsnm{Yabuno}, \binits{M.}},
\bauthor{\bsnm{China}, \binits{F.}},
\bauthor{\bsnm{Terai}, \binits{H.}},
\bauthor{\bsnm{Furusawa}, \binits{A.}}:
\batitle{{Quantum detector tomography of a superconducting nanostrip photon-number-resolving detector}}.
\bjtitle{Optics Express}
\bvolume{29}(\bissue{8}),
\bfpage{11728}
(\byear{2021})
\doiurl{10.1364/OE.423142}
{\href{https://arxiv.org/abs/2102.09712}{{2102.09712}}}
\end{barticle}
\endbibitem

\bibitem[\protect\citeauthoryear{Cai et~al.}{2021}]{cai2021quantum}
\begin{barticle}
\bauthor{\bsnm{Cai}, \binits{Y.}},
\bauthor{\bsnm{Chen}, \binits{Y.}},
\bauthor{\bsnm{Chen}, \binits{X.}},
\bauthor{\bsnm{Wu}, \binits{G.}},
\bauthor{\bsnm{Wu}, \binits{E.}}:
\batitle{{Quantum characteristics and applications of multi‐pixel photon counter}}.
\bjtitle{Microwave and Optical Technology Letters}
\bvolume{63}(\bissue{8}),
\bfpage{2052}--\blpage{2057}
(\byear{2021})
\doiurl{10.1002/mop.32865}
\end{barticle}
\endbibitem

\bibitem[\protect\citeauthoryear{Fitzke et~al.}{2022}]{fitzke2022time}
\begin{barticle}
\bauthor{\bsnm{Fitzke}, \binits{E.}},
\bauthor{\bsnm{Krebs}, \binits{R.}},
\bauthor{\bsnm{Haase}, \binits{T.}},
\bauthor{\bsnm{Mengler}, \binits{M.}},
\bauthor{\bsnm{Alber}, \binits{G.}},
\bauthor{\bsnm{Walther}, \binits{T.}}:
\batitle{{Time-dependent POVM reconstruction for single-photon avalanche photo diodes using adaptive regularization}}.
\bjtitle{New Journal of Physics}
(\byear{2022})
\doiurl{10.1088/1367-2630/ac5004}
\end{barticle}
\endbibitem

\bibitem[\protect\citeauthoryear{Santana et~al.}{2023}]{santana2023extending_preprint}
\begin{botherref}
\oauthor{\bsnm{Santana}, \binits{T.}},
\oauthor{\bsnm{Mu{\~{n}}oz}, \binits{C.}},
\oauthor{\bsnm{Chunnilall}, \binits{C.}}:
{Extending the quantum tomography of a quasi-photon-number-resolving detector}
(2023)
\doiurl{10.1364/opticaopen.24908667.v1}
\end{botherref}
\endbibitem

\bibitem[\protect\citeauthoryear{Cooper et~al.}{2014}]{cooper2014local}
\begin{barticle}
\bauthor{\bsnm{Cooper}, \binits{M.}},
\bauthor{\bsnm{Karpi{\'{n}}ski}, \binits{M.}},
\bauthor{\bsnm{Smith}, \binits{B.J.}}:
\batitle{{Local mapping of detector response for reliable quantum state estimation}}.
\bjtitle{Nature Communications}
\bvolume{5}(\bissue{1}),
\bfpage{4332}
(\byear{2014})
\doiurl{10.1038/ncomms5332}
\end{barticle}
\endbibitem

\bibitem[\protect\citeauthoryear{Schapeler et~al.}{2021}]{schapeler2021quantum}
\begin{barticle}
\bauthor{\bsnm{Schapeler}, \binits{T.}},
\bauthor{\bsnm{H{\"{o}}pker}, \binits{J.P.}},
\bauthor{\bsnm{Bartley}, \binits{T.J.}}:
\batitle{{Quantum detector tomography of a high dynamic-range superconducting nanowire single-photon detector}}.
\bjtitle{Superconductor Science and Technology}
\bvolume{34}(\bissue{6}),
\bfpage{64002}
(\byear{2021})
\doiurl{10.1088/1361-6668/abee9a}
\end{barticle}
\endbibitem

\bibitem[\protect\citeauthoryear{Liu et~al.}{2023}]{liu2023optimized}
\begin{barticle}
\bauthor{\bsnm{Liu}, \binits{D.-S.}},
\bauthor{\bsnm{Wang}, \binits{J.-Q.}},
\bauthor{\bsnm{Zou}, \binits{C.-L.}},
\bauthor{\bsnm{Ren}, \binits{X.-F.}},
\bauthor{\bsnm{Guo}, \binits{G.-C.}}:
\batitle{{Optimized detector tomography for photon-number-resolving detectors with hundreds of pixels}}.
\bjtitle{Physical Review A}
\bvolume{108}(\bissue{5}),
\bfpage{052611}
(\byear{2023})
\doiurl{10.1103/PhysRevA.108.052611}
\end{barticle}
\endbibitem

\bibitem[\protect\citeauthoryear{Zhang et~al.}{2012}]{zhang2012mapping}
\begin{barticle}
\bauthor{\bsnm{Zhang}, \binits{L.}},
\bauthor{\bsnm{Coldenstrodt-Ronge}, \binits{H.B.}},
\bauthor{\bsnm{Datta}, \binits{A.}},
\bauthor{\bsnm{Puentes}, \binits{G.}},
\bauthor{\bsnm{Lundeen}, \binits{J.S.}},
\bauthor{\bsnm{Jin}, \binits{X.-M.}},
\bauthor{\bsnm{Smith}, \binits{B.J.}},
\bauthor{\bsnm{Plenio}, \binits{M.B.}},
\bauthor{\bsnm{Walmsley}, \binits{I.A.}}:
\batitle{{Mapping coherence in measurement via full quantum tomography of a hybrid optical detector}}.
\bjtitle{Nature Photonics}
\bvolume{6}(\bissue{6}),
\bfpage{364}--\blpage{368}
(\byear{2012})
\doiurl{10.1038/nphoton.2012.107}
\end{barticle}
\endbibitem

\bibitem[\protect\citeauthoryear{Morimoto et~al.}{2020}]{morimoto2020megapixel}
\begin{barticle}
\bauthor{\bsnm{Morimoto}, \binits{K.}},
\bauthor{\bsnm{Ardelean}, \binits{A.}},
\bauthor{\bsnm{Wu}, \binits{M.-L.}},
\bauthor{\bsnm{Ulku}, \binits{A.C.}},
\bauthor{\bsnm{Antolovic}, \binits{I.M.}},
\bauthor{\bsnm{Bruschini}, \binits{C.}},
\bauthor{\bsnm{Charbon}, \binits{E.}}:
\batitle{{Megapixel time-gated SPAD image sensor for 2D and 3D imaging applications}}.
\bjtitle{Optica}
\bvolume{7}(\bissue{4}),
\bfpage{346}
(\byear{2020})
\doiurl{10.1364/OPTICA.386574}
\end{barticle}
\endbibitem

\bibitem[\protect\citeauthoryear{{MOSEK ApS}}{2023}]{mosek}
\begin{botherref}
\oauthor{\bsnm{{MOSEK ApS}}}:
The MOSEK Optimization Toolbox for MATLAB Manual. Version 10.1.
(2023).
\url{http://docs.mosek.com/10.1/toolbox/index.html}
\end{botherref}
\endbibitem

\bibitem[\protect\citeauthoryear{Diamond and Boyd}{2016}]{diamond2016cvxpy}
\begin{barticle}
\bauthor{\bsnm{Diamond}, \binits{S.}},
\bauthor{\bsnm{Boyd}, \binits{S.}}:
\batitle{{CVXPY}: {A} {P}ython-embedded modeling language for convex optimization}.
\bjtitle{Journal of Machine Learning Research}
\bvolume{17}(\bissue{83}),
\bfpage{1}--\blpage{5}
(\byear{2016})
\end{barticle}
\endbibitem

\bibitem[\protect\citeauthoryear{Agrawal et~al.}{2018}]{agrawal2018rewriting}
\begin{barticle}
\bauthor{\bsnm{Agrawal}, \binits{A.}},
\bauthor{\bsnm{Verschueren}, \binits{R.}},
\bauthor{\bsnm{Diamond}, \binits{S.}},
\bauthor{\bsnm{Boyd}, \binits{S.}}:
\batitle{{A rewriting system for convex optimization problems}}.
\bjtitle{Journal of Control and Decision}
\bvolume{5}(\bissue{1}),
\bfpage{42}--\blpage{60}
(\byear{2018})
\doiurl{10.1080/23307706.2017.1397554}
\end{barticle}
\endbibitem

\bibitem[\protect\citeauthoryear{Bauer et~al.}{2024}]{noctua2}
\begin{botherref}
\oauthor{\bsnm{Bauer}, \binits{C.}},
\oauthor{\bsnm{Kenter}, \binits{T.}},
\oauthor{\bsnm{Lass}, \binits{M.}},
\oauthor{\bsnm{Mazur}, \binits{L.}},
\oauthor{\bsnm{Meyer}, \binits{M.}},
\oauthor{\bsnm{Nitsche}, \binits{H.}},
\oauthor{\bsnm{Riebler}, \binits{H.}},
\oauthor{\bsnm{Schade}, \binits{R.}},
\oauthor{\bsnm{Schwarz}, \binits{M.}},
\oauthor{\bsnm{Winnwa}, \binits{N.}},
\oauthor{\bsnm{Wiens}, \binits{A.}},
\oauthor{\bsnm{Wu}, \binits{X.}},
\oauthor{\bsnm{Plessl}, \binits{C.}},
\oauthor{\bsnm{Simon}, \binits{J.}}:
Noctua 2 supercomputer.
Journal of large-scale research facilities JLSRF
(2024).
In press.
\end{botherref}
\endbibitem

\bibitem[\protect\citeauthoryear{Bertsekas}{1982}]{bertsekas1982}
\begin{barticle}
\bauthor{\bsnm{Bertsekas}, \binits{D.P.}}:
\batitle{Projected newton methods for optimization problems with simple constraints}.
\bjtitle{SIAM Journal on Control and Optimization}
\bvolume{20}(\bissue{2}),
\bfpage{221}--\blpage{246}
(\byear{1982})
\doiurl{10.1137/0320018}
{\href{https://arxiv.org/abs/https://doi.org/10.1137/0320018}{{https://doi.org/10.1137/0320018}}}
\end{barticle}
\endbibitem

\bibitem[\protect\citeauthoryear{Landi and {Loli Piccolomini}}{2008}]{landi2008a}
\begin{barticle}
\bauthor{\bsnm{Landi}, \binits{G.}},
\bauthor{\bsnm{{Loli Piccolomini}}, \binits{E.}}:
\batitle{{A projected Newton-CG method for nonnegative astronomical image deblurring}}.
\bjtitle{Numerical Algorithms}
\bvolume{48}(\bissue{4}),
\bfpage{279}--\blpage{300}
(\byear{2008})
\doiurl{10.1007/s11075-008-9198-3}
\end{barticle}
\endbibitem

\bibitem[\protect\citeauthoryear{Schmidt et~al.}{2011}]{schmidt2011optimization}
\begin{botherref}
\oauthor{\bsnm{Schmidt}, \binits{M.}},
\oauthor{\bsnm{Kim}, \binits{D.}},
\oauthor{\bsnm{Sra}, \binits{S.}}:
{Projected Newton-type Methods in Machine Learning}.
The MIT Press
(2011).
\doiurl{10.7551/mitpress/8996.003.0013}
\end{botherref}
\endbibitem

\bibitem[\protect\citeauthoryear{Tiedau et~al.}{2019}]{tiedau2019high}
\begin{barticle}
\bauthor{\bsnm{Tiedau}, \binits{J.}},
\bauthor{\bsnm{Meyer-Scott}, \binits{E.}},
\bauthor{\bsnm{Nitsche}, \binits{T.}},
\bauthor{\bsnm{Barkhofen}, \binits{S.}},
\bauthor{\bsnm{Bartley}, \binits{T.J.}},
\bauthor{\bsnm{Silberhorn}, \binits{C.}}:
\batitle{{A high dynamic range optical detector for measuring single photons and bright light}}.
\bjtitle{Optics Express}
\bvolume{27}(\bissue{1}),
\bfpage{1}--\blpage{15}
(\byear{2019})
\doiurl{10.1364/OE.27.000001}
\end{barticle}
\endbibitem

\bibitem[\protect\citeauthoryear{Liu et~al.}{2023}]{Liugithub}
\begin{botherref}
\oauthor{\bsnm{Liu}, \binits{D.-S.}},
\oauthor{\bsnm{Wang}, \binits{J.-Q.}},
\oauthor{\bsnm{Zou}, \binits{C.-L.}},
\oauthor{\bsnm{Ren}, \binits{X.-F.}},
\oauthor{\bsnm{Guo}, \binits{G.-C.}}:
Optimized detector tomography for photon-number resolving detectors with hundreds of pixels
(2023).
\url{https://github.com/DS-Liu/Modified-detector-tomography}
\end{botherref}
\endbibitem

\bibitem[\protect\citeauthoryear{Cheng et~al.}{2023}]{cheng2023a}
\begin{barticle}
\bauthor{\bsnm{Cheng}, \binits{R.}},
\bauthor{\bsnm{Zhou}, \binits{Y.}},
\bauthor{\bsnm{Wang}, \binits{S.}},
\bauthor{\bsnm{Shen}, \binits{M.}},
\bauthor{\bsnm{Taher}, \binits{T.}},
\bauthor{\bsnm{Tang}, \binits{H.X.}}:
\batitle{{A 100-pixel photon-number-resolving detector unveiling photon statistics}}.
\bjtitle{Nature Photonics}
\bvolume{17}(\bissue{1}),
\bfpage{112}--\blpage{119}
(\byear{2023})
\doiurl{10.1038/s41566-022-01119-3}
\end{barticle}
\endbibitem

\bibitem[\protect\citeauthoryear{Eaton et~al.}{2023}]{eaton2023resolution}
\begin{barticle}
\bauthor{\bsnm{Eaton}, \binits{M.}},
\bauthor{\bsnm{Hossameldin}, \binits{A.}},
\bauthor{\bsnm{Birrittella}, \binits{R.J.}},
\bauthor{\bsnm{Alsing}, \binits{P.M.}},
\bauthor{\bsnm{Gerry}, \binits{C.C.}},
\bauthor{\bsnm{Dong}, \binits{H.}},
\bauthor{\bsnm{Cuevas}, \binits{C.}},
\bauthor{\bsnm{Pfister}, \binits{O.}}:
\batitle{{Resolution of 100 photons and quantum generation of unbiased random numbers}}.
\bjtitle{Nature Photonics}
\bvolume{17}(\bissue{1}),
\bfpage{106}--\blpage{111}
(\byear{2023})
\doiurl{10.1038/s41566-022-01105-9}
\end{barticle}
\endbibitem

\bibitem[\protect\citeauthoryear{Schade et~al.}{2024}]{pqdts}
\begin{botherref}
\oauthor{\bsnm{Schade}, \binits{R.}},
\oauthor{\bsnm{Lass}, \binits{M.}},
\oauthor{\bsnm{Schapeler}, \binits{T.}},
\oauthor{\bsnm{Plessl}, \binits{C.}},
\oauthor{\bsnm{Bartley}, \binits{T.J.}}:
Parallel Quantum Detector Tomography Solver (pqdts).
\doiurl{10.5281/zenodo.10908474} .
\url{https://github.com/pc2/pqdts}
\end{botherref}
\endbibitem

\bibitem[\protect\citeauthoryear{Fernandez and Williams}{2010}]{fernandez2010closed}
\begin{barticle}
\bauthor{\bsnm{Fernandez}, \binits{M.}},
\bauthor{\bsnm{Williams}, \binits{S.}}:
\batitle{{Closed-Form Expression for the Poisson-Binomial Probability Density Function}}.
\bjtitle{IEEE Transactions on Aerospace and Electronic Systems}
\bvolume{46}(\bissue{2}),
\bfpage{803}--\blpage{817}
(\byear{2010})
\doiurl{10.1109/TAES.2010.5461658}
\end{barticle}
\endbibitem

\bibitem[\protect\citeauthoryear{Powell}{1969}]{powellmethod}
\begin{botherref}
\oauthor{\bsnm{Powell}, \binits{M.J.}}:
{A method for nonlinear constraints in minimization problems}.
Optimization,
283--298
(1969)
\end{botherref}
\endbibitem

\bibitem[\protect\citeauthoryear{Hestenes}{1969}]{Hestenes1969}
\begin{barticle}
\bauthor{\bsnm{Hestenes}, \binits{M.R.}}:
\batitle{Multiplier and gradient methods}.
\bjtitle{J. Optim. Theory Appl.}
\bvolume{4}(\bissue{5}),
\bfpage{303}--\blpage{320}
(\byear{1969})
\doiurl{10.1007/BF00927673}
\end{barticle}
\endbibitem

\bibitem[\protect\citeauthoryear{Nocedal and Wright}{2006}]{NoceWrig06}
\begin{bbook}
\bauthor{\bsnm{Nocedal}, \binits{J.}},
\bauthor{\bsnm{Wright}, \binits{S.J.}}:
\bbtitle{Numerical Optimization},
\bedition{2e} edn.
\bpublisher{Springer},
\blocation{New York, NY, USA}
(\byear{2006})
\end{bbook}
\endbibitem

\bibitem[\protect\citeauthoryear{Condat}{2016}]{Condat2016}
\begin{barticle}
\bauthor{\bsnm{Condat}, \binits{L.}}:
\batitle{Fast projection onto the simplex and the l1 ball}.
\bjtitle{Mathematical Programming}
\bvolume{158}(\bissue{1}),
\bfpage{575}--\blpage{585}
(\byear{2016})
\doiurl{10.1007/s10107-015-0946-6}
\end{barticle}
\endbibitem

\bibitem[\protect\citeauthoryear{{Message Passing Interface Forum}}{2023}]{mpi41}
\begin{botherref}
\oauthor{\bsnm{{Message Passing Interface Forum}}}:
{MPI}: A Message-Passing Interface Standard Version 4.1.
(2023).
\url{https://www.mpi-forum.org/docs/mpi-4.1/mpi41-report.pdf}
\end{botherref}
\endbibitem

\bibitem[\protect\citeauthoryear{Dagum and Menon}{1998}]{openmp}
\begin{barticle}
\bauthor{\bsnm{Dagum}, \binits{L.}},
\bauthor{\bsnm{Menon}, \binits{R.}}:
\batitle{{OpenMP}: An industry-standard {API} for shared-memory programming}.
\bjtitle{IEEE Comput. Sci. Eng.}
\bvolume{5}(\bissue{1}),
\bfpage{46}--\blpage{55}
(\byear{1998})
\doiurl{10.1109/99.660313}
\end{barticle}
\endbibitem

\bibitem[\protect\citeauthoryear{Zhang et~al.}{2012}]{zhang2012recursive}
\begin{barticle}
\bauthor{\bsnm{Zhang}, \binits{L.}},
\bauthor{\bsnm{Datta}, \binits{A.}},
\bauthor{\bsnm{Coldenstrodt-Ronge}, \binits{H.B.}},
\bauthor{\bsnm{Jin}, \binits{X.-M.}},
\bauthor{\bsnm{Eisert}, \binits{J.}},
\bauthor{\bsnm{Plenio}, \binits{M.B.}},
\bauthor{\bsnm{Walmsley}, \binits{I.A.}}:
\batitle{{Recursive quantum detector tomography}}.
\bjtitle{New Journal of Physics}
\bvolume{14}(\bissue{11}),
\bfpage{115005}
(\byear{2012})
\doiurl{10.1088/1367-2630/14/11/115005}
\end{barticle}
\endbibitem

\bibitem[\protect\citeauthoryear{Chen et~al.}{2022}]{chen2022efficient}
\begin{barticle}
\bauthor{\bsnm{Chen}, \binits{X.}},
\bauthor{\bsnm{Xu}, \binits{F.}},
\bauthor{\bsnm{Xu}, \binits{H.}},
\bauthor{\bsnm{Zhang}, \binits{L.}}:
\batitle{{Efficient tomography of coherent optical detectors}}.
\bjtitle{Physical Review A}
\bvolume{106}(\bissue{5}),
\bfpage{051702}
(\byear{2022})
\doiurl{10.1103/PhysRevA.106.L051702}
\end{barticle}
\endbibitem

\bibitem[\protect\citeauthoryear{Higham}{1988}]{higham1988computing}
\begin{barticle}
\bauthor{\bsnm{Higham}, \binits{N.J.}}:
\batitle{{Computing a nearest symmetric positive semidefinite matrix}}.
\bjtitle{Linear Algebra and its Applications}
\bvolume{103},
\bfpage{103}--\blpage{118}
(\byear{1988})
\doiurl{10.1016/0024-3795(88)90223-6}
\end{barticle}
\endbibitem

\bibitem[\protect\citeauthoryear{Francisco and Gon{\c{c}}alves}{2017}]{francisco2017a}
\begin{barticle}
\bauthor{\bsnm{Francisco}, \binits{J.B.}},
\bauthor{\bsnm{Gon{\c{c}}alves}, \binits{D.S.}}:
\batitle{{A fixed-point method for approximate projection onto the positive semidefinite cone}}.
\bjtitle{Linear Algebra and its Applications}
\bvolume{523},
\bfpage{59}--\blpage{78}
(\byear{2017})
\doiurl{10.1016/j.laa.2017.02.014}
\end{barticle}
\endbibitem

\bibitem[\protect\citeauthoryear{Kr{\"{a}}mer et~al.}{2018}]{kramer2018quantumoptics}
\begin{barticle}
\bauthor{\bsnm{Kr{\"{a}}mer}, \binits{S.}},
\bauthor{\bsnm{Plankensteiner}, \binits{D.}},
\bauthor{\bsnm{Ostermann}, \binits{L.}},
\bauthor{\bsnm{Ritsch}, \binits{H.}}:
\batitle{{QuantumOptics.jl: A Julia framework for simulating open quantum systems}}.
\bjtitle{Computer Physics Communications}
\bvolume{227},
\bfpage{109}--\blpage{116}
(\byear{2018})
\doiurl{10.1016/j.cpc.2018.02.004}
\end{barticle}
\endbibitem

\bibitem[\protect\citeauthoryear{Fousse et~al.}{2007}]{fousse2007mpfr}
\begin{barticle}
\bauthor{\bsnm{Fousse}, \binits{L.}},
\bauthor{\bsnm{Hanrot}, \binits{G.}},
\bauthor{\bsnm{Lef\`{e}vre}, \binits{V.}},
\bauthor{\bsnm{P\'{e}lissier}, \binits{P.}},
\bauthor{\bsnm{Zimmermann}, \binits{P.}}:
\batitle{Mpfr: A multiple-precision binary floating-point library with correct rounding}.
\bjtitle{ACM Trans. Math. Softw.}
\bvolume{33}(\bissue{2}),
\bfpage{13}
(\byear{2007})
\doiurl{10.1145/1236463.1236468}
\end{barticle}
\endbibitem

\end{thebibliography}

\end{document}